\documentclass[preprint2]{aastex61}

\usepackage{graphicx}
\usepackage{epstopdf}
\usepackage{mathptmx} 
\usepackage{natbib}

\DeclareSymbolFont{matha}{OML}{txmi}{m}{it}
\DeclareMathSymbol{\varv}{\mathord}{matha}{118}

\newcommand{\Lsun}{L$_{\odot}$}

\newcommand{\msec}[2]{$#1\mbox{$''\mskip-7.6mu.\,$}#2$}
\newcommand{\mdeg}[2]{$#1\mbox{$^\circ\mskip-7.6mu.\,$}#2$}

\newcommand{\iras}{IRAS~16293--2422}
\newcommand{\sbeam}[7]{\msec{#1}{#2}$\times$\msec{#3}{#4} ; $#5$\mdeg{#6}{#7}}

\received{XXX}
\revised{XXX}
\accepted{XXX}

\shorttitle{On the nature of the compact sources in \iras}
\shortauthors{}

\begin{document}

\title{On the nature of the compact sources in \iras\ seen in at centimeter to sub-millimeter wavelengths}


\author{Antonio Hern\'andez-G\'omez}
\affiliation{Instituto de Radioastronom\'{\i}a y Astrof\'{\i}sica, Universidad Nacional Aut\'onoma de M\'exico, 58089 Morelia, Mexico}
\affiliation{IRAP, Universit\'e de Toulouse, CNRS, UPS, CNES, Toulouse, France}

\author{Laurent Loinard}
\affiliation{Instituto de Radioastronom\'{\i}a y Astrof\'{\i}sica, Universidad Nacional Aut\'onoma de M\'exico, 58089 Morelia, Mexico}
\affiliation{Instituto de Astronom\'{\i}a, Universidad Nacional Aut\'onoma de M\'exico, Apartado Postal 70-264, 04510 Cuidad de M\'exico, Mexico}

\author{Claire J.\ Chandler}
\affiliation{National Radio Astronomy Observatory, 1003 Lopezville Rd, Socorro, NM 87801, USA}

\author{Luis F.\ Rodr\'{\i}guez}
\affiliation{Instituto de Radioastronom\'{\i}a y Astrof\'{\i}sica, Universidad Nacional Aut\'onoma de M\'exico, 58089 Morelia, Mexico}

\author{Luis A.\ Zapata}
\affiliation{Instituto de Radioastronom\'{\i}a y Astrof\'{\i}sica, Universidad Nacional Aut\'onoma de M\'exico, 58089 Morelia, Mexico}

\author{David J.\ Wilner}
\affiliation{Harvard-Smithsonian Center for Astrophysics, 60 Garden Street, Cambridge, MA 02138, USA}

\author{Paul T.P.\ Ho}
\affiliation{Institute of Astronomy and Astrophysics, Academia Sinica, P.O. Box 23-141, Taipei 10617, Taiwan}
\affiliation{East Asian Observatory, 660 N. Aohoku Place University Park, Hilo, Hawaii 96720, USA}

\author{Emmanuel Caux}
\affiliation{IRAP, Universit\'e de Toulouse, CNRS, UPS, CNES, Toulouse, France}

\author{David Qu\'enard}
\affiliation{School of Physics and Astronomy, Queen Mary University of London, Mile End Road, London E1 4NS, UK}

\author{Sandrine Bottinelli}
\affiliation{IRAP, Universit\'e de Toulouse, CNRS, UPS, CNES, Toulouse, France}

\author{Crystal L.\ Brogan}
\affiliation{National Radio Astronomy Observatory, 520 Edgemont Rd, Charlottesville, VA 22903, USA}

\author{Lee Hartmann}
\affiliation{Department of Astronomy, University of Michigan, 1085 S. University Ave., Ann Arbor, MI 48109, USA}

\author{Karl M. Menten}
\affiliation{Max-Planck-Institut f\"ur Radioastronomie, Auf dem H\"ugel 69, D-53121 Bonn, Germany }

\begin{abstract}
We present multi-epoch continuum observations of the Class 0 protostellar system \iras\ taken with the Very Large Array (VLA) at multiple wavelengths between 7 mm and 15 cm (41 GHz down to 2 GHz), as well as single-epoch Atacama Large Millimeter/submillimeter Array (ALMA) continuum observations covering the range from 0.4 to 1.3 mm (700 GHz down to 230 GHz). The new VLA observations confirm that source A2 is a protostar driving episodic mass ejections, and reveal the complex relative motion between  A2 and A1. The spectrum of component B can be described by a single power law ($S_\nu \propto \nu^{2.28}$) over the entire range from 3 to 700 GHz (10 cm down to 0.4 mm), suggesting that the emission is entirely dominated by dust even at $\lambda$ = 10 cm. Finally, the size of source B appears to increase with frequency up to 41 GHz, remaining roughly constant (at \msec{0}{39} $\equiv$ 55 AU) at higher frequencies. We interpret this as evidence that source B is a dusty structure of finite size that becomes increasingly optically thick at higher frequencies until, in the millimeter regime, the source becomes entirely optically thick. The lack of excess free-free emission at long wavelengths, combined with the absence of high-velocity molecular emission indicates that source B does not drive a powerful outflow, and might indicate that source B is at a particularly early stage of its evolution.
\end{abstract}

\keywords{ISM: individual (\iras) --- star: formation --- ISM: jets and outflows --- astrometry --- binaries: visual --- techniques: interferometric}

\section{Introduction \label{sect:intro} }

A paradigm exists for the formation of isolated low-mass stars through the gravitational collapse of dense cores embedded in a large molecular cloud \citep{Shu1987}. The different evolutionary stages considered in this scenario, from Class 0 to Class I, Flat Spectrum, and Class II and III sources, have been amply described in the literature \citep{Lada1984,Andre1993,Greene1994}. Although many features of this paradigm can be extended to low-mass stars forming in multiple systems, several standing questions remain regarding the formation and early evolution of multiple systems. For instance, the dominant route(s) leading to the formation of multiple stellar systems are still debated. The two leading (and not necessarily mutually exclusive) contending theories are turbulent and disk fragmentation \citep{Padoan2007,Adams1989}. The turbulent fragmentation theory considers that a bound core can break into multiple fragments due to turbulent fluctuations of the density. These fragments will have masses larger that the Jean mass and will collapse faster than the original core. In this scenario, the binary or multiple systems are expected to form if these fragments remain gravitationally bound \citep[see e.g.][]{goodwin2004}. On the other hand, the disk fragmentation model considers that strong gravitational instabilities can fragment a pre-existing disk to form a multiple system \citep{Adams1989}. Both theories make different predictions on the architecture of the resulting multiple systems that can be tested through high spatial resolution observations \citep{Tobin2016a,Tobin2016b}. For instance, determining parameters from the observations such as the distance between the companions, their relative orientation or eccentricity might help us distinguish between these scenarios.

 Furthermore, multiplicity directly affects early stellar evolution through a number of tidal mechanisms \citep[e.g.][]{Arty1994,Kraus2011} that have been discussed in detail by \citet{Reipurth2014} and that can be particularly severe in young triple systems. These effects can, in particular, affect the accretion history and time evolution of members of multiple systems, rendering the interpretation of their observational properties in terms of theoretical models somewhat uncertain \citep[e.g.][]{Stassun2008}. To study them, young multiple systems must be characterized in detail and at high angular resolution. Very young (Class~0) systems are particularly interesting in this context, because they probe the initial conditions of multiple stellar evolution. 

\iras\ is a very well known Class 0 protostellar system located in the Lynds~1689N dark cloud within the Ophiuchus star-forming region. For some time, the distance to the entire Ophiuchus complex was assumed to be 120 pc following \citet{Loinard2008}. In a recent work, \citet{OrtizLeon2017} determined trigonometric parallaxes to a sample of young stars distributed over the region and inferred mean distances of 137.3 $\pm$ 1.2 pc for the Ophiuchus core, Lynds~1688, and 147.3 $\pm$ 3.4 pc for Lynds~1689. \citet{Dzib2018}, using astrometric observations of water masers, recently confirmed that the distance to \iras\ is $141_{-21}^{+30}$ pc, which is consistent with both \citet{Loinard2008} and \citet{OrtizLeon2017}, so we will use 141 pc for the distance to \iras\ in the rest of the paper. 

\begin{figure*}[h!]
\centering
\setlength\tabcolsep{0.001pt}
\begin{tabular}{c c c}

\includegraphics[width=0.33\textwidth,trim= 0 0 3 5,clip]{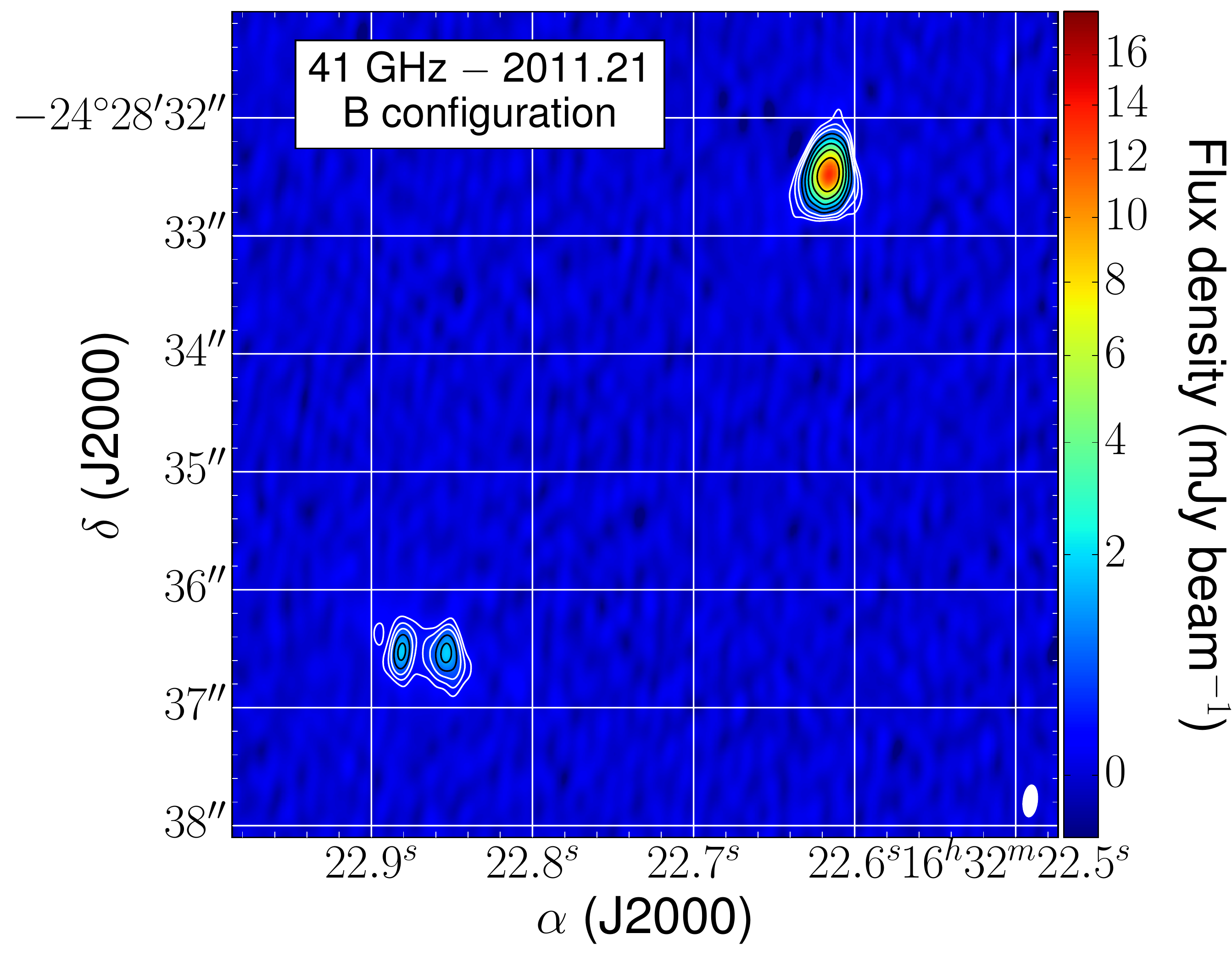} &
\includegraphics[width=0.32\textwidth,trim= 0 0 3 5,clip]{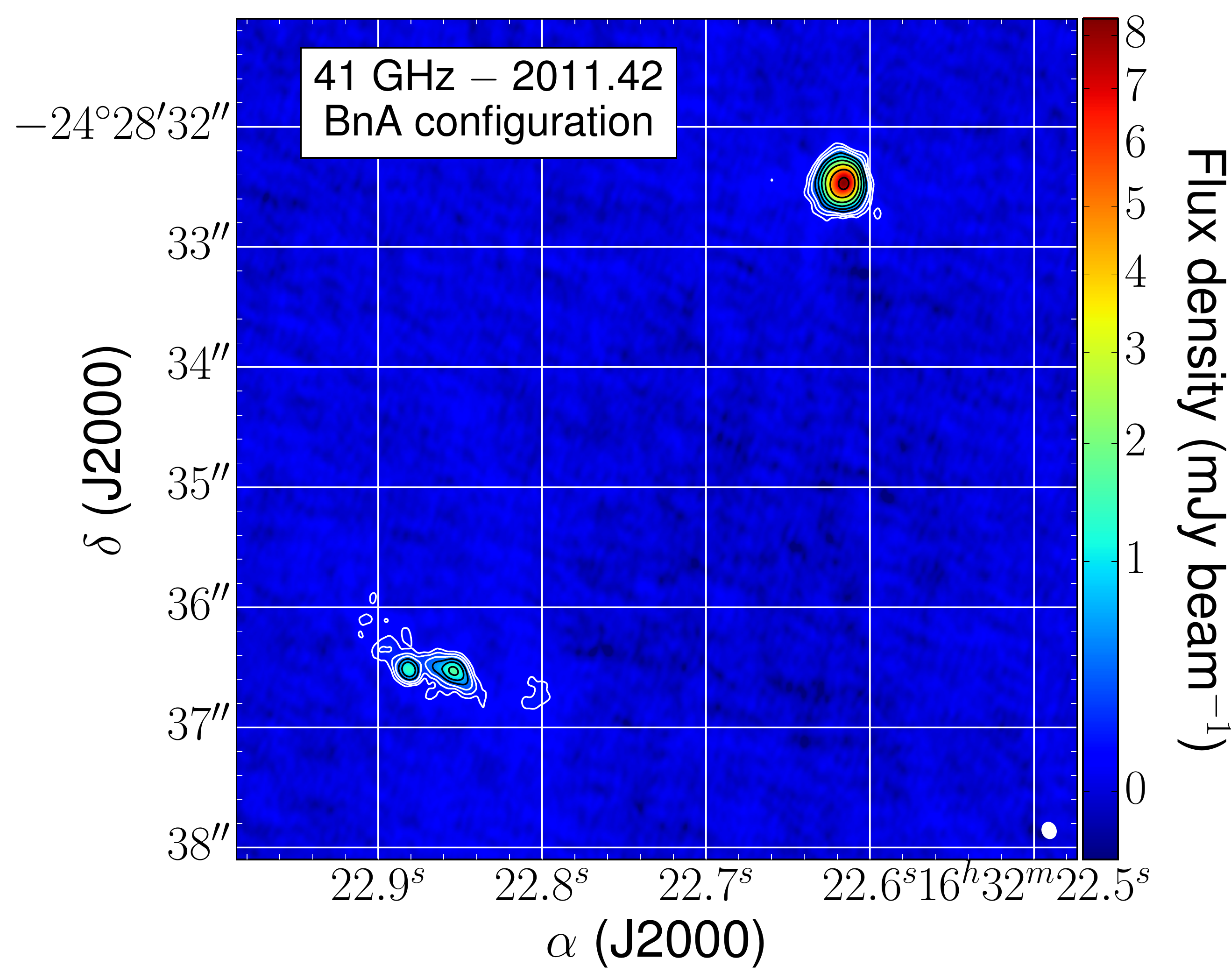} &  
\includegraphics[width=0.335\textwidth,trim= 0 0 3 5,clip]{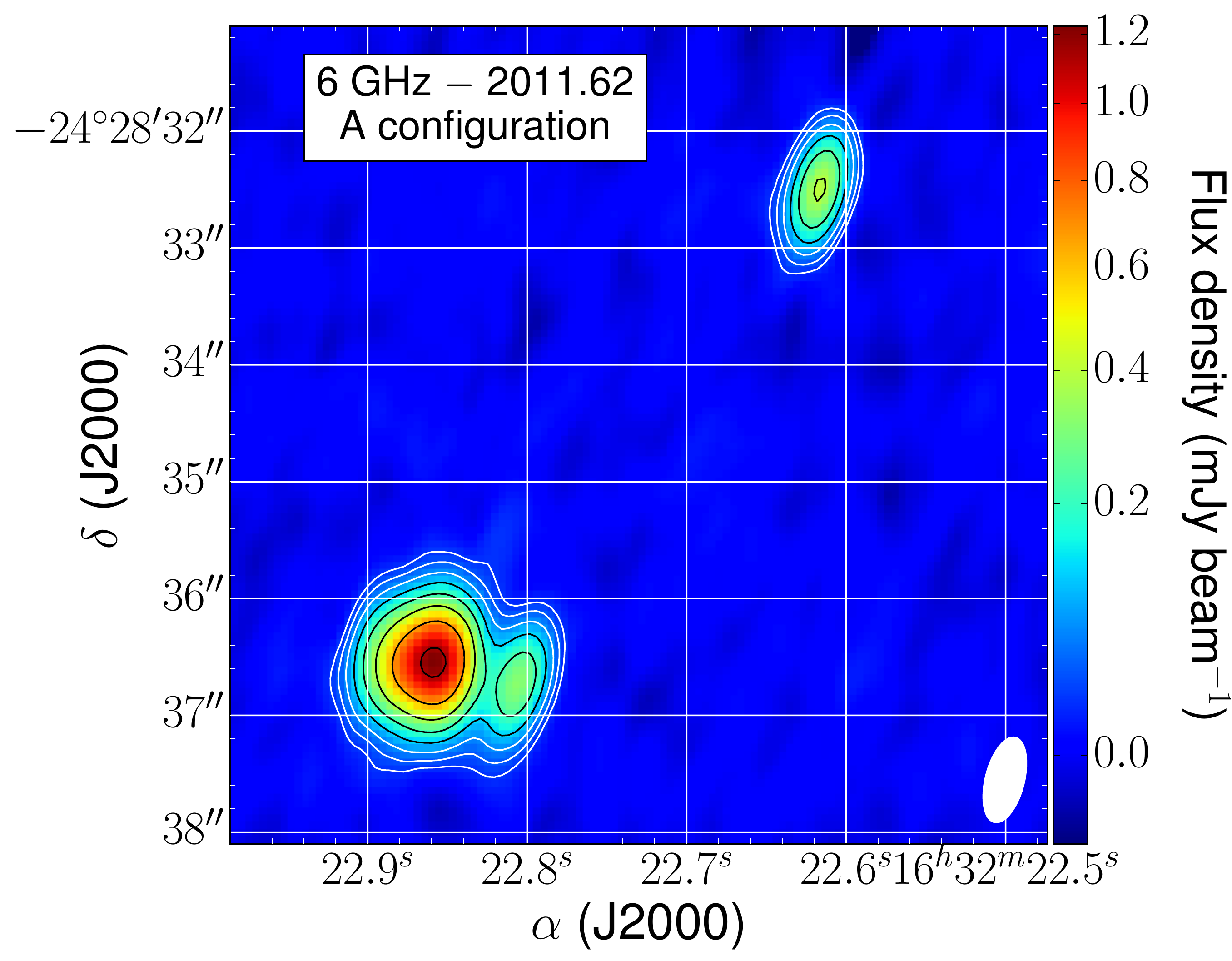} \\

\includegraphics[width=0.33\textwidth,trim= 0 0 3 5,clip]{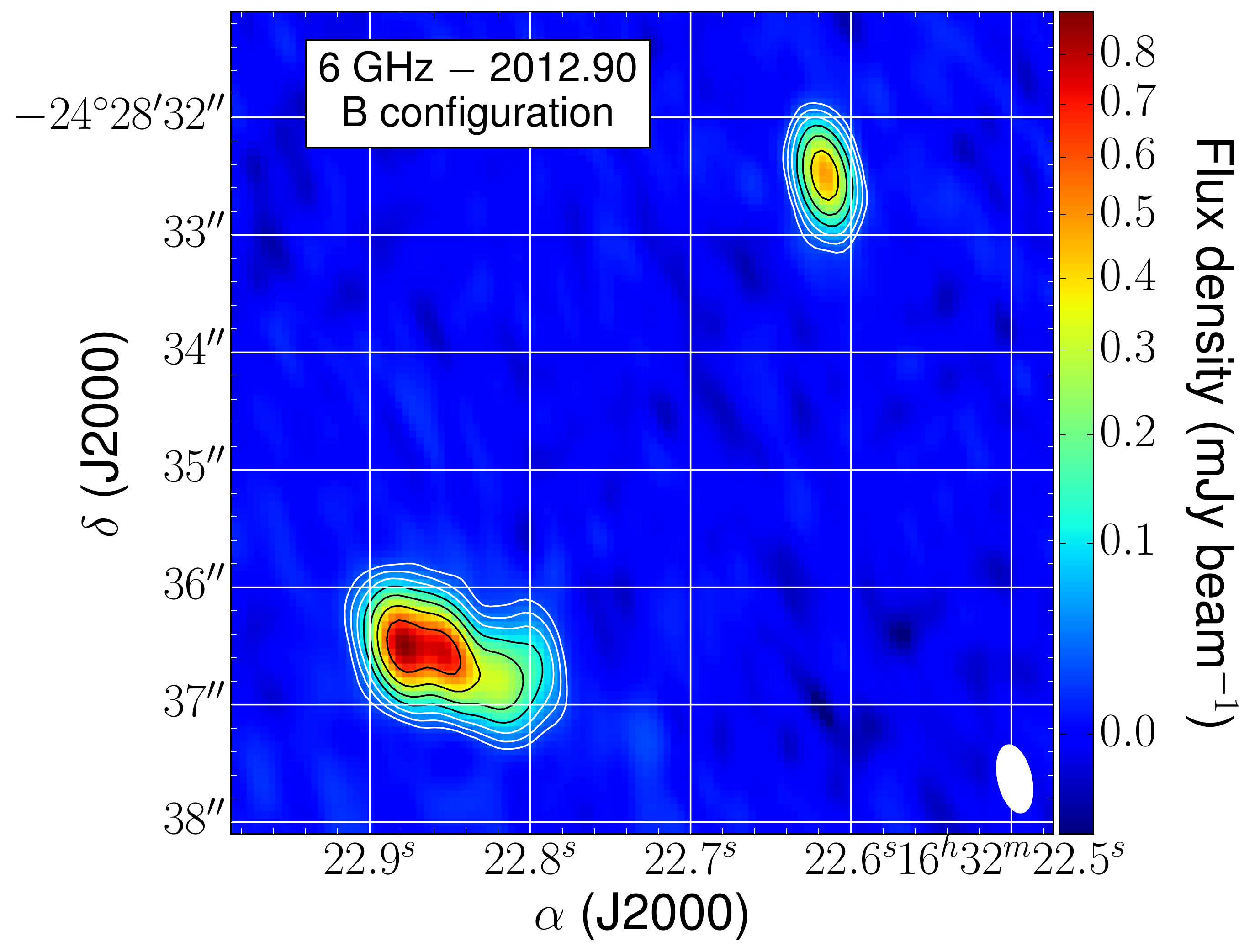} &
\includegraphics[width=0.33\textwidth,trim= 0 0 3 5,clip]{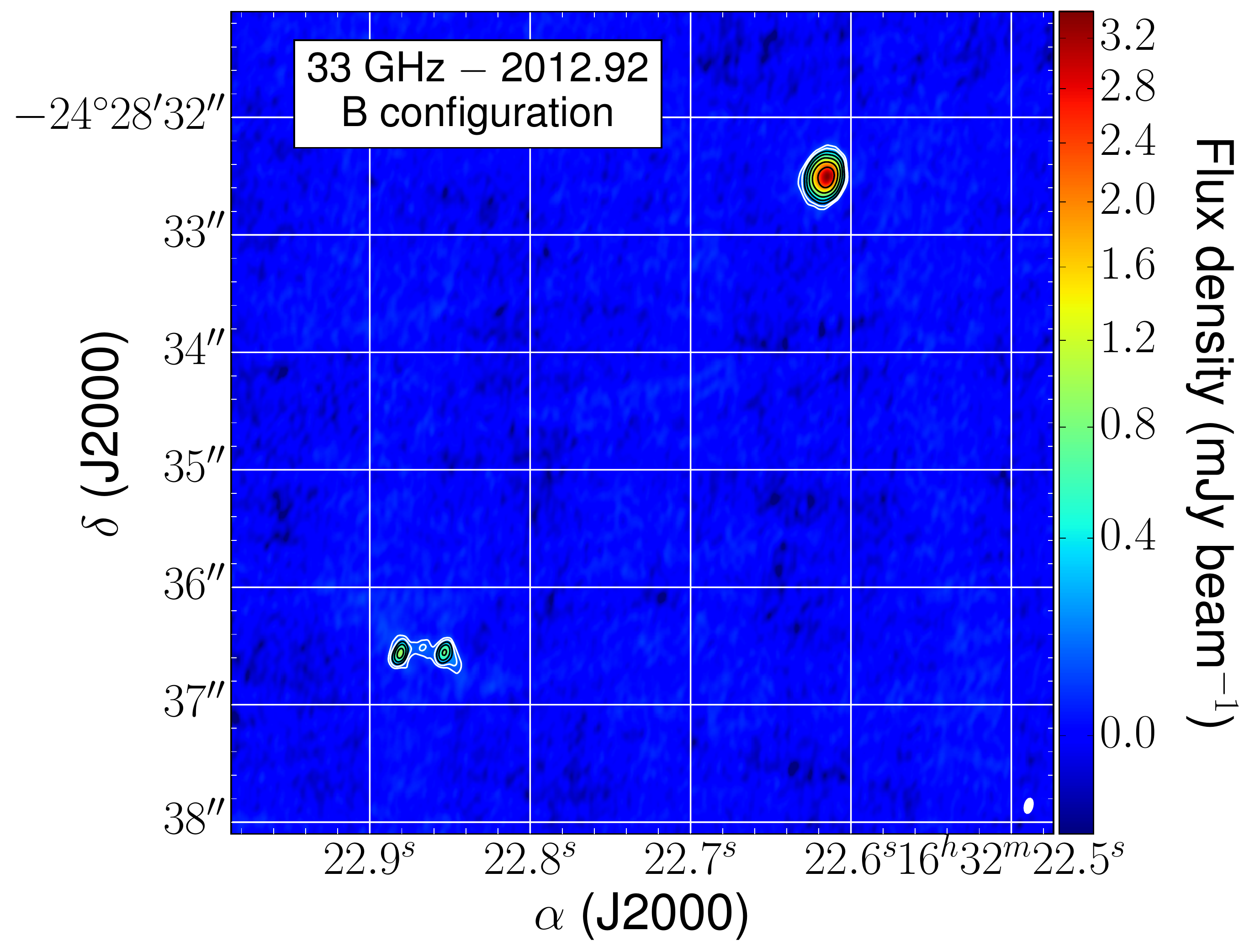} & 
\includegraphics[width=0.33\textwidth,trim= 0 0 3 5,clip]{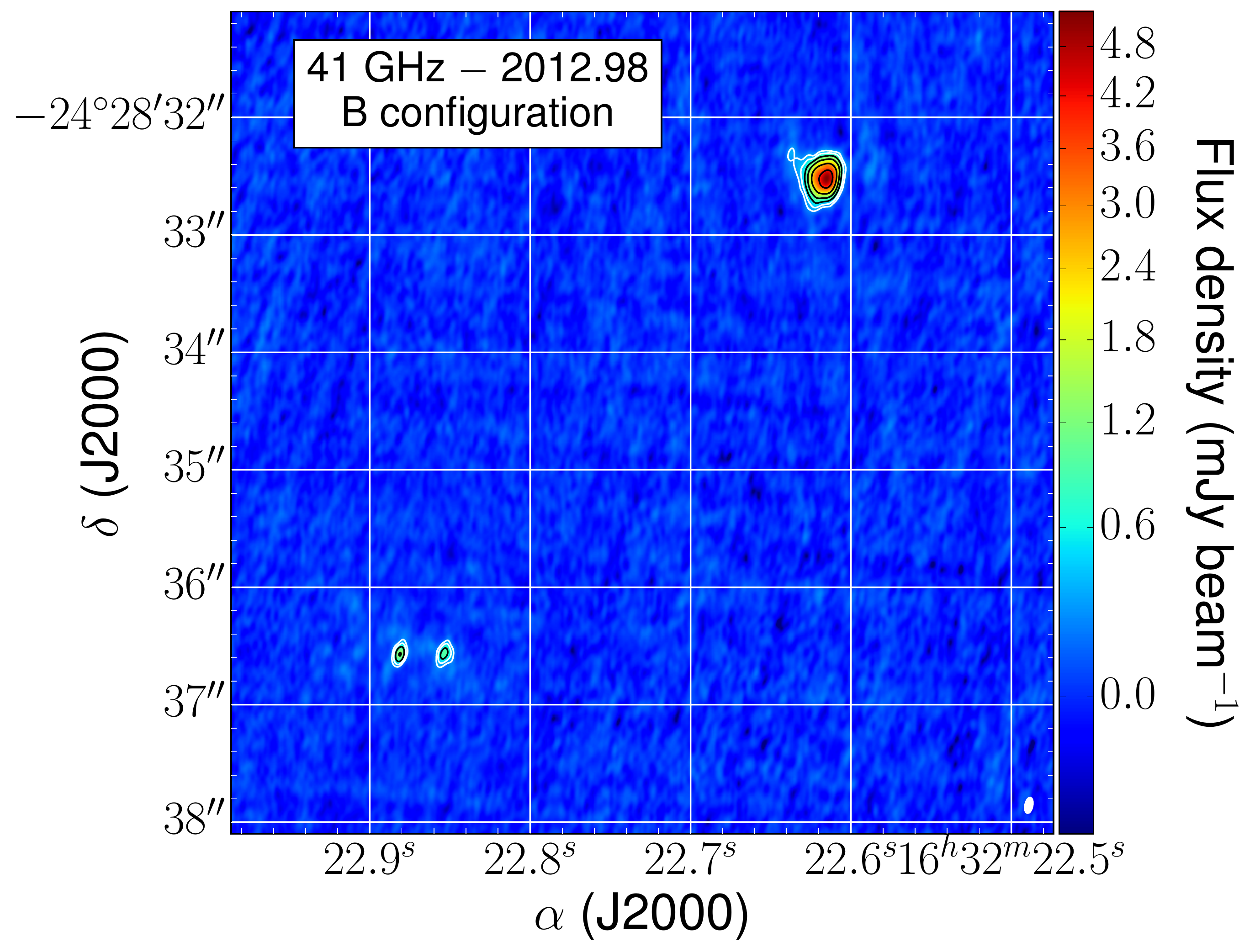} \\

\includegraphics[width=0.33\textwidth,trim= 0 0 3 5, clip]{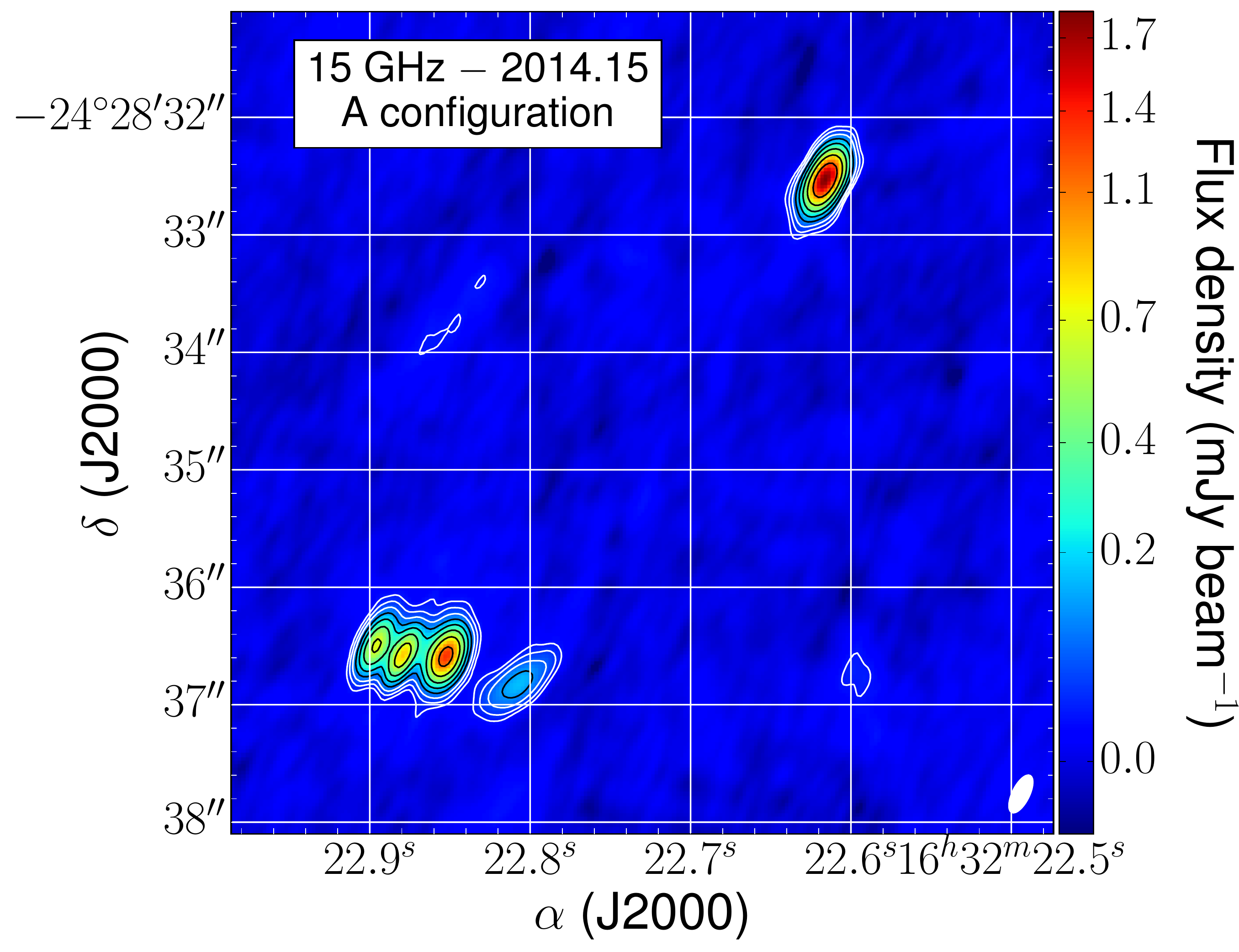} & 
\includegraphics[width=0.33\textwidth,trim= 0 0 3 5,clip]{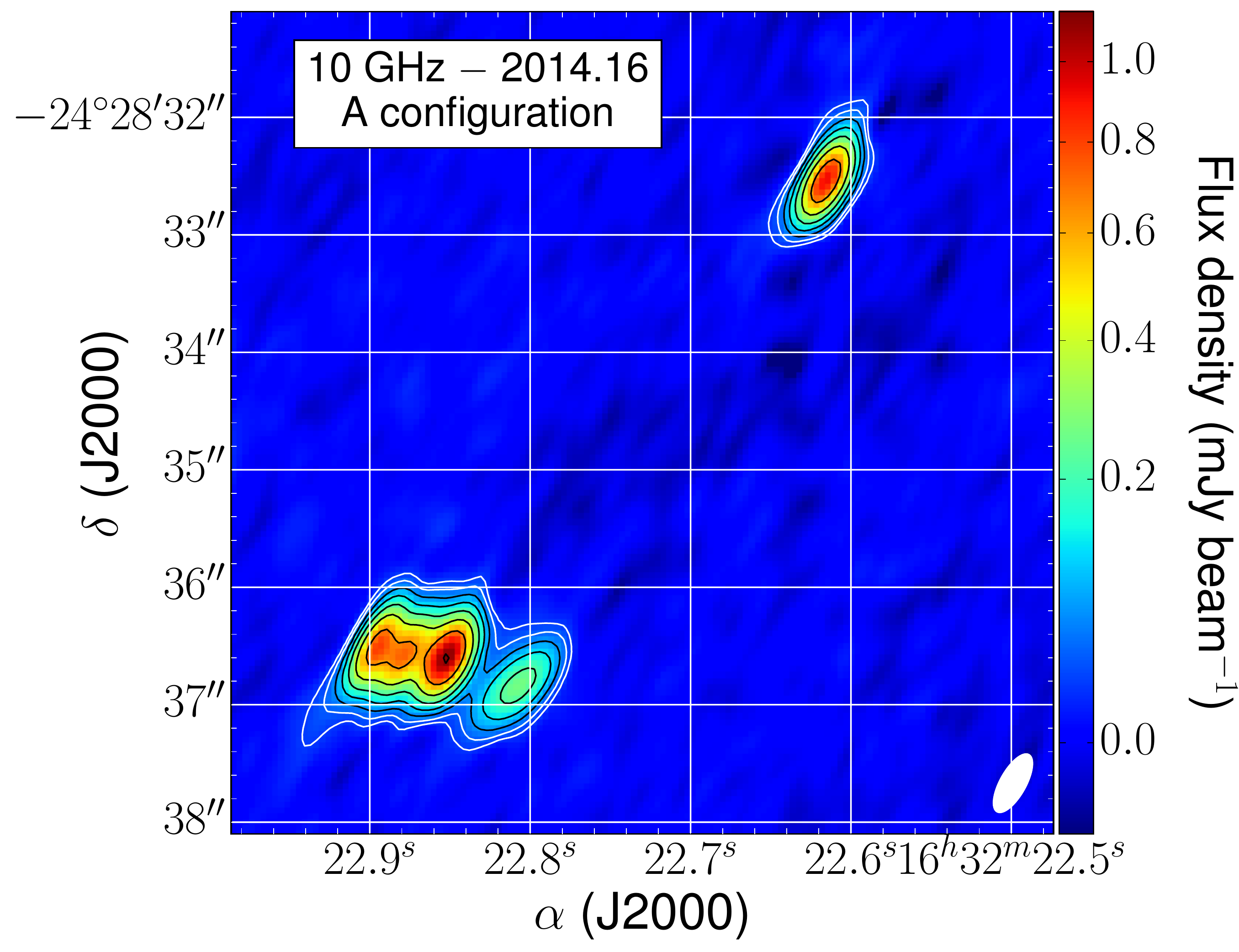} & 
\includegraphics[width=0.33\textwidth,trim= 0 0 3 5, clip]{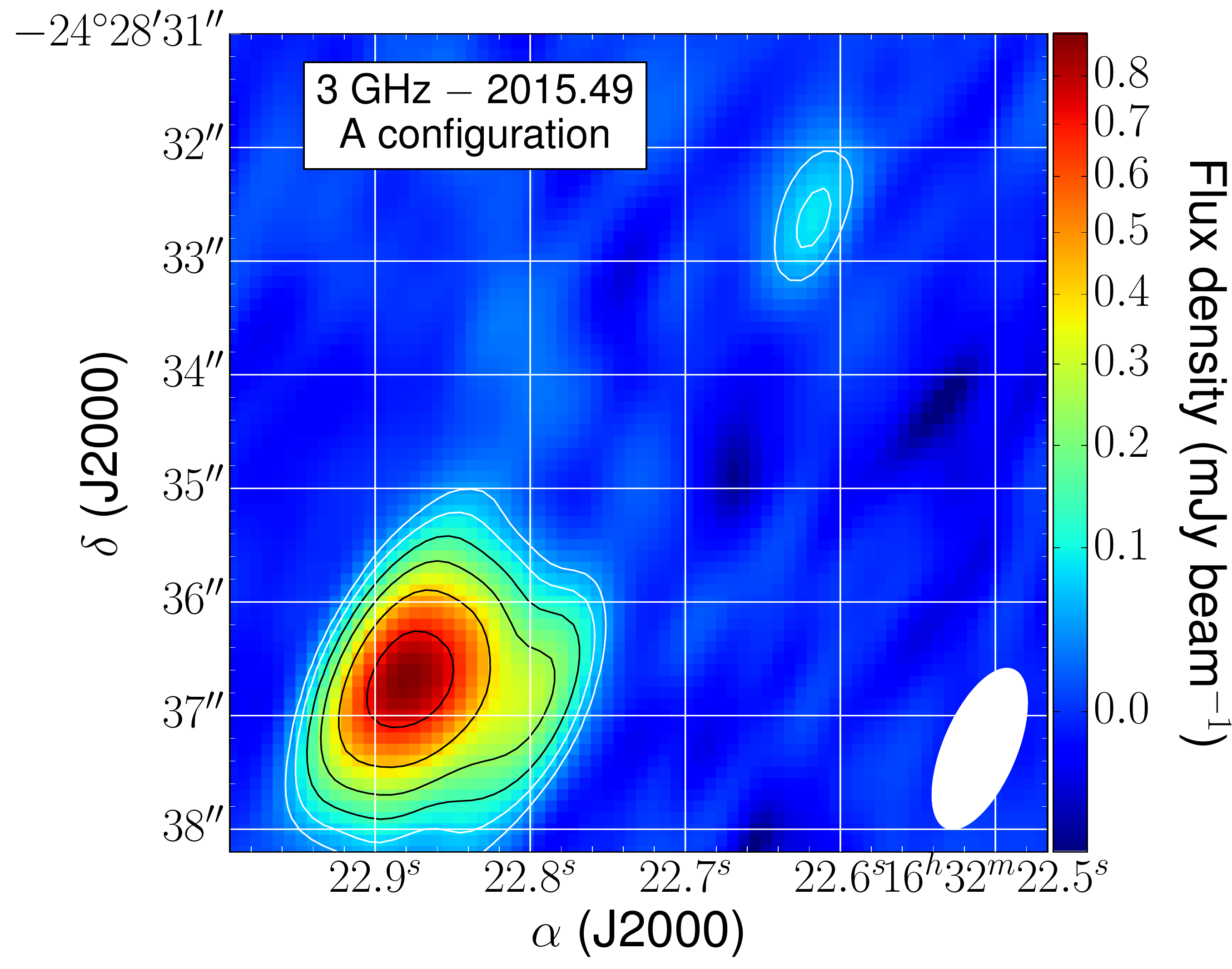} \\

\includegraphics[width=0.33\textwidth,trim= 0 0 3 5,clip]{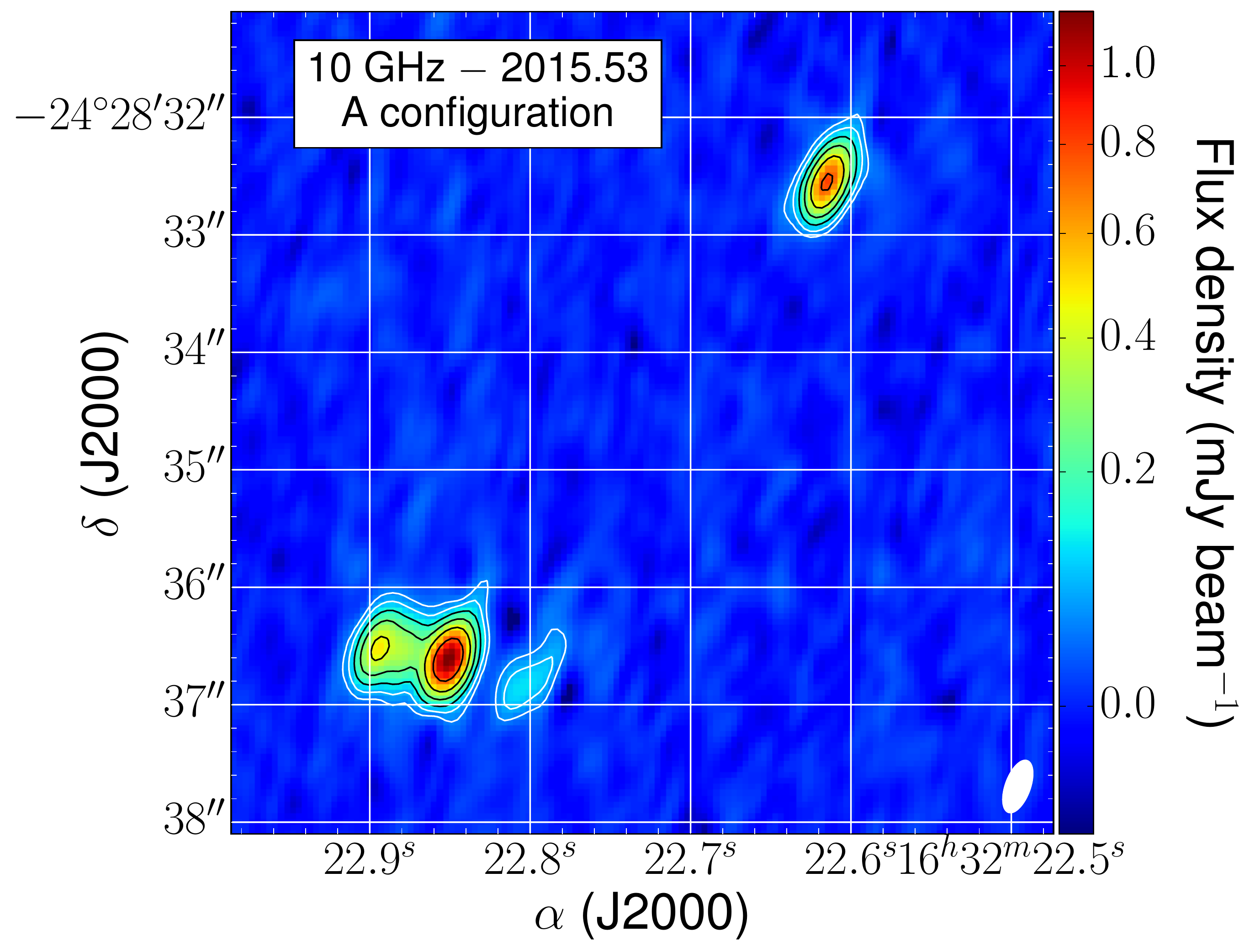} &  
\includegraphics[width=0.33\textwidth,trim= 0 0 3 5, clip]{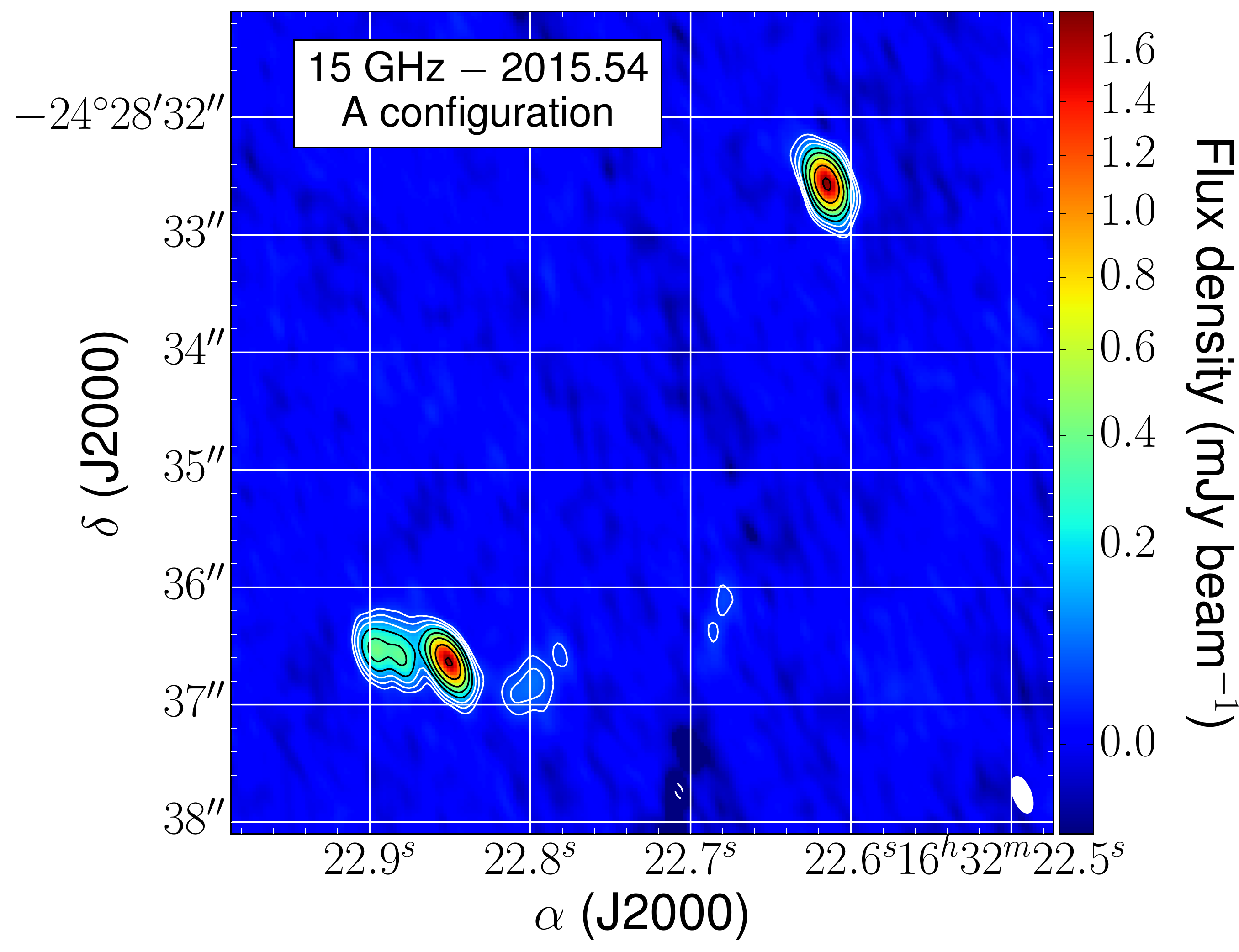}  \\

\end{tabular}
\caption{Continuum images of \iras\ for all new VLA observations presented in this paper. For all of them, the first contour corresponds to five times the r.m.s.\ noise of each image, and successive contours increase by a factor $\sqrt{3}$. The two main components, A and B, are located, respectively, to the south-east and north-west of the images (see also Figure \ref{fig:zoom}). }
\label{fig:all_images}
\end{figure*}

\begin{figure*}[!t]
\centering
\setlength\tabcolsep{0.001pt}
\begin{tabular}{c c}
\includegraphics[angle=0,width=0.668\textwidth,trim = 0 0 5 5,clip]{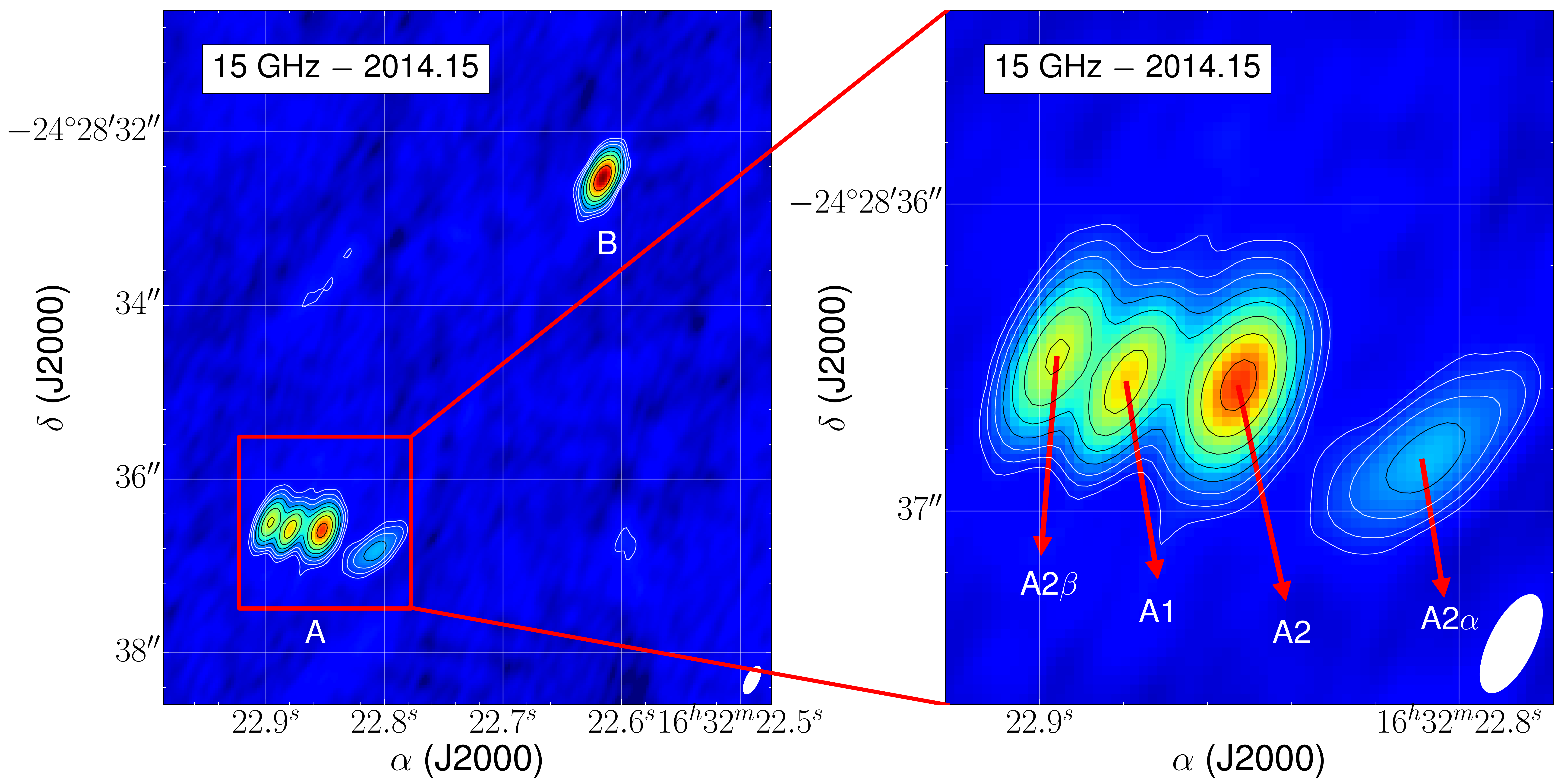}  &
\includegraphics[angle=0,width=0.333\textwidth,trim = 0 0 5 5,clip]{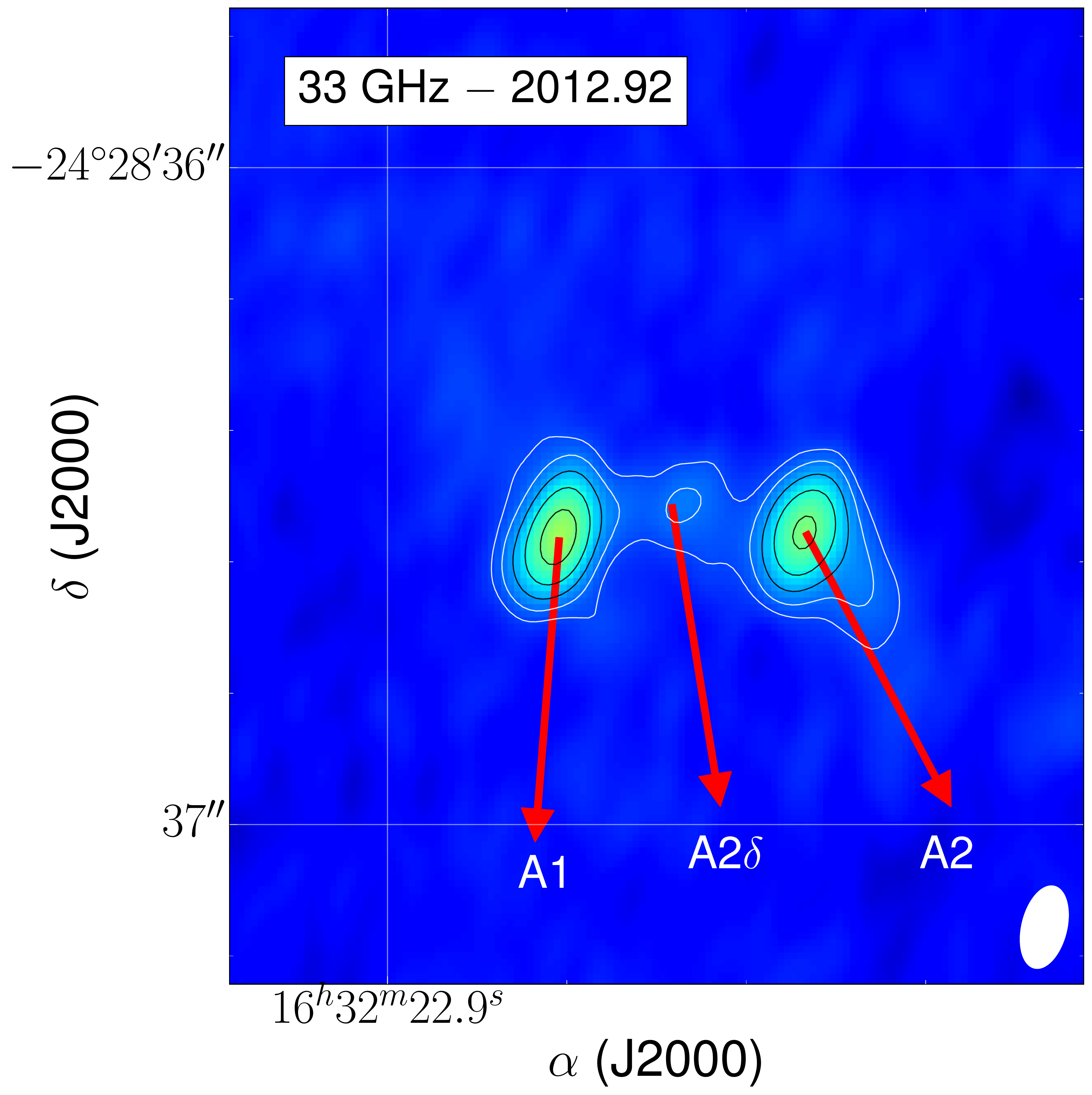}
\end{tabular}
\caption{Labelling of the various sources in \iras. Left panel: VLA image at 15 GHz indicating the position of sources A and B. Middle panel: zoom on source A at 15 GHz. The sources A1 and A2 are indicated as well as the ejecta A2$\alpha$, A2$\beta$ (see text). Right panel: VLA image at 33 GHz. The sources A1, A2 and A2$\delta$ are labeled. The contours and the color scale corresponding to the flux density are the same as in Figure \ref{fig:all_images}.}
\label{fig:zoom}
\end{figure*}

\iras\ has been amply studied over the years for a variety of reasons. Initially, it drew attention because it was the coldest known protostar \citep[see e.g.][]{walker1986}. Later on, it was found that this source presented very rich spectra with numerous complex molecules, indicating a particularly active chemistry. Indeed, it harbors the archetypical ``hot corino'' at its center \citep[e.g.][]{Ceccarelli2000,Caux2011,Jorgensen2016}. Finally, \iras\ happens to be one of the first very young (Class~0) multiple stellar systems ever identified. Interferometric observations at centimeter and millimeter wavelengths by \citet{Wootten1989} and \citet{Mundy1992} revealed two compact sources (called A and B) near the center of the extended envelope of \iras. At about the same time, \citet{Mizuno1990} identified two compact outflows driven from \iras, confirming that it must contain a young multiple system. High resolution radio observations \citep{Wootten1989,Loinard2002} have further revealed that source A contains two sub-condensations called A1 and A2. Multi-epoch radio observations \citep{Loinard2002,Chandler2005,Loinard2007,Pech2010} have shown that, while the separation between A1 and A2 remained constant at about \msec{0}{34} between 1986 and 2008, their relative position angle monotonically increased by 40$^\circ$ over the same time period. \citet{Loinard2007} and \citet{Pech2010} interpreted this relative motion in terms of a nearly face-on, nearly circular orbit, and therefore considered A1 and A2 to form a tight binary system (see Figures \ref{fig:all_images} and \ref{fig:zoom}). 

One of the two compact outflows identified by \citet{Mizuno1990} is oriented almost exactly in the E--W direction, while the other is oriented in the NE--SW direction (at a position angle of about 65$^\circ$). High resolution observations by \citet{Yeh2008} demonstrated unambiguously that the E--W outflow originates from within source A. On the other hand, source A2 was shown by \citet{Loinard2007} to be at the origin of multiple bipolar ejections giving rise to additional radio sources (called A2$\alpha$, A2$\beta$, etc.) oriented along the same direction as the NE--SW ouflow. \citet{Pech2010} used multi-epoch radio observations to show that A2$\alpha$ and A2$\beta$ are symmetrically moving away from A2, as expected for bipolar ejecta. Thus, the NE--SW outflow can be unambiguously traced back to A2, and both compact outflows identified by \citet{Mizuno1990} appear to be driven from within source A. A third compact outflow (oriented in the NW--SE direction) was identified in \iras\ by \citet{Rao2009} and studied in more detail by \citet{Girart2014}. In their interpretation, this outflow also originates from within source A and impinges on source B where it splits, producing an arc-like structure. \citet{Loinard2013} offered a different interpretation in terms of a slow and poorly collimated outflow driven by source B. Regardless of the origin of this third outflow, it appears abundantly clear that source A drives multiple outflows and must host a very young multiple system.

\begin{deluxetable*}{lrclccc}
\tablecaption{Parameters of the interferometric observations used in this paper. \label{tab:obs}}
\tablewidth{0pt}
\tablehead{
\colhead{Project} & \colhead{Date}               & \colhead{Frequency} & \colhead{Synthesized beam}                                                    &\colhead{r.m.s. noise}        & \colhead{S$_A$}           & \colhead{S$_B$}\\
\colhead{}        & \colhead{\sc (dd/mm/yyyy)}   & \colhead{(GHz)}     &\colhead{($\theta_{\textrm{max}} \times \theta_{\textrm{min}}$ ; P.A.)}          & \colhead{(mJy bm$^{-1})$} & \colhead{(mJy)} & \colhead{(mJy)}}
\startdata
\multicolumn{7}{l}{\bf VLA Data:}\\
\decimals
15A-363  & 28/06/2015        &  3.0 & \sbeam{1}{10}{0}{46}{-}{20}{7}  & 0.013  & 1.70 $\pm$ 0.17 & 0.052 $\pm$ 0.005 \\
15A-363  & 15/07/2015        & 10.0 & \sbeam{0}{35}{0}{14}{-}{17}{5}  & 0.014  & 2.95 $\pm$ 0.30 & 1.08  $\pm$ 0.11  \\
15A-363  & 17/07/2015        & 15.0 & \sbeam{0}{23}{0}{10}{+}{19}{6}  & 0.008  & 3.84 $\pm$ 0.38 & 2.64  $\pm$ 0.26  \\
14A-313  & 25/02/2014        & 15.0 & \sbeam{0}{24}{0}{09}{-}{23}{7}  & 0.005  & 4.39 $\pm$ 0.44 & 2.68  $\pm$ 0.27  \\
14A-313  & 27/02/2014        & 10.0 & \sbeam{0}{39}{0}{14}{-}{26}{9}  & 0.006  & 3.61 $\pm$ 0.36 & 1.05  $\pm$ 0.11  \\
12B-088  & 25/11/2012        &  7.0 & \sbeam{0}{44}{0}{19}{-}{168}{4} & 0.006  & 2.58 $\pm$ 0.26 & 0.55  $\pm$ 0.06  \\
12B-088  & 01/12/2012        & 33.0 & \sbeam{0}{11}{0}{04}{-}{15}{6}  & 0.027  & 3.63 $\pm$ 0.54 & 15.3  $\pm$ 2.3   \\
12B-088  & 23/12/2012        & 41.0 & \sbeam{0}{09}{0}{04}{-}{4}{6}   & 0.072  & 4.87 $\pm$ 0.73 & 26.2  $\pm$ 3.9   \\
10C-222  & 15/03/2011        & 41.0 & \sbeam{0}{36}{0}{16}{-}{10}{0}  & 0.030  & 6.59 $\pm$ 0.99 & 31.3  $\pm$ 4.7   \\
10C-222  & 19/03/2011        & 41.0 & \sbeam{0}{30}{0}{14}{+}{11}{5}  & 0.060  & 6.86 $\pm$ 1.03 & 28.9  $\pm$ 4.3   \\
10C-222  & 05/06/2011        & 41.0 & \sbeam{0}{13}{0}{10}{-}{13}{9}  & 0.027  & 5.27 $\pm$ 0.79 & 26.9  $\pm$ 4.0   \\ 
10C-222  & 08/06/2011        & 41.0 & \sbeam{0}{08}{0}{05}{-}{174}{4} & 0.037  & 5.81 $\pm$ 0.87 & 29.1  $\pm$ 4.4   \\
10C-222  & 13/08/2011        &  6.0 & \sbeam{0}{50}{0}{20}{-}{12}{5}  & 0.006  & 2.72 $\pm$ 0.27 & 0.38  $\pm$ 0.04  \\
\hline
\multicolumn{5}{l}{\bf ALMA Data:}\\
01005762 & 17/08/2014	    & 227.0 	& \sbeam{0}{53}{0}{25}{+}{87}{2} & 1.81  & 1.47 $\pm$ 0.22 & 1.79 $\pm$ 0.27 \\
01003905 & 14/06/2014	    & 318.0 	& \sbeam{0}{39}{0}{34}{-}{62}{2} & 3.89  & 3.81 $\pm$ 0.57 & 3.21 $\pm$ 0.48 \\
01003908 & 14/06/2014	    & 318.0 	& \sbeam{0}{35}{0}{31}{-}{62}{2} & 2.27  & 3.92 $\pm$ 0.59 & 3.23 $\pm$ 0.48 \\
01003968 & 16/06/2014	    & 323.0 	& \sbeam{0}{39}{0}{32}{+}{87}{3} & 3.15  & 3.79 $\pm$ 0.57 & 3.35 $\pm$ 0.50 \\
01028453 & 28/06/2015	    & 338.2 	& \sbeam{0}{17}{0}{13}{-}{83}{9} & 3.49  & 2.46 $\pm$ 0.37 & 3.60 $\pm$ 0.54 \\
01028026 & 29/08/2015	    & 342.3 	& \sbeam{0}{25}{0}{13}{-}{76}{9} & 6.03  & 2.78 $\pm$ 0.42 & 3.12 $\pm$ 0.57 \\
01019938 & 06/06/2015	    & 404.0 	& \sbeam{0}{28}{0}{22}{-}{78}{3} & 7.95  & 6.06 $\pm$ 1.21 & 4.67 $\pm$ 0.93 \\
01019922 & 07/06/2015	    & 453.0 	& \sbeam{0}{31}{0}{21}{-}{81}{9} & 8.86  & 8.86 $\pm$ 1.77 & 6.14 $\pm$ 1.23 \\
SV       & 17/04/2012       & 695.0     & \sbeam{0}{29}{0}{16}{-}{70}{4} & 19.90 & 11.9 $\pm$ 2.4  & 13.5 $\pm$ 2.7  \\
\enddata
\tablecomments{\scriptsize Observational details for VLA data obtained before 2011 can be found in \citet{Chandler2005}, \citet{Loinard2007}, and \citet{Pech2010}. The ALMA project numbers correspond to the ID numbers given in the Japanese Virtual Observatory (JVO) from where the data sets were obtained. For the ALMA observation at 695 GHz, SV stands for Science Verification.}
\label{tab:obs}
\end{deluxetable*}

Source B in \iras\ has also been studied in detail. Although it is resolved and does exhibit some substructure when observed at very high angular resolution \citep[e.g.][]{Rodriguez2005}, there is no evidence to suggest that it might harbour a multiple system. Source B exhibits two remarkable properties that are worth mentioning here. The first one is that its spectrum can be accurately described as a single power law, with a spectral index of order 2 to 2.5 from $\nu$ = 5 GHz ($\lambda$ = 6 cm) to $\nu$ = 330 GHz ($\lambda$ = 0.8 mm) \citep{Chandler2005}. This suggests that a single emission mechanism, thermal dust emission, is at work over that entire frequency range. This is highly unusual, since emission from low-mass protostars at centimeter wavelengths is almost always dominated by free-free radiation from an ionized wind \citep[e.g.][]{Anglada2015}. The interpretation of the centimeter flux from source B in terms of dust emission was further confirmed by the observation by \citet{Chandler2005} that the size of source B {\em increases} with frequency in the centimeter regime as expected for optically thick thermal dust emission. In contrast, optically thick free-free emission is expected to result in a source size that {\em decreases} with frequency \citep[e.g.][]{Panagia1975}. The lack of free-free emission in source B would be expected in the scenario put forward by \citet{Girart2014} where all molecular outflow activity in \iras\ is driven from source A. We note, however, that it would not necessarily be incompatible with the slow and poorly collimated outflow proposed by \citet{Loinard2013} which might not generate the highly supersonic speeds required to produce shock-ionized gas.
 
The second remarkable feature of source B worth mentioning here is that it exhibits what is arguably the clearest known example of an inverse P-Cygni profile \citep{Pineda2012,Zapata2013}. This is naturally interpreted as evidence for infall, and \citet{Pineda2012} derived a mass accretion rate on source B of $4.5\times 10^{-5}$ M$_{\odot}$ yr$^{-1}$. \citet{Zapata2013} confirmed the interpretation of the P-Cygni profiles in terms of infall, and interpret source B itself as an optically thick disk seen nearly face-on. The high accretion rate onto source B is somewhat difficult to reconcile with the lack of a strong wind inferred from the absence of free-free emission, and of outflowing material proposed by \citet{Girart2014}. Indeed, theoretical models of accreting protostars predict a mass ratio between accreted and ejected material of order 10 to 30\% \citep[e.g.][]{Shu1988}.

It is clear from the description presented above that \iras\ is in the process of forming a multiple (at least triple) stellar system. As such, it offers a unique opportunity to investigate the effect of multiplicity on the earliest stages of stellar evolution. Nevertheless, the exact number of protostars within \iras, their relative evolutionary stages, and their relationship with known radio sources and outflows in the system, remain elusive. In this paper, we present new radio, millimeter, and sub-millimeter continuum observations aimed at elucidating several of these aspects. The observations will be presented in Section 2, the results given in Section 3, and used to discuss the nature of the sources in Section 4. Section 5 summarizes our conclusions and provides some perspectives.

\begin{figure*}[!t]
\centering
\setlength\tabcolsep{0.001pt}
\includegraphics[scale=0.42]{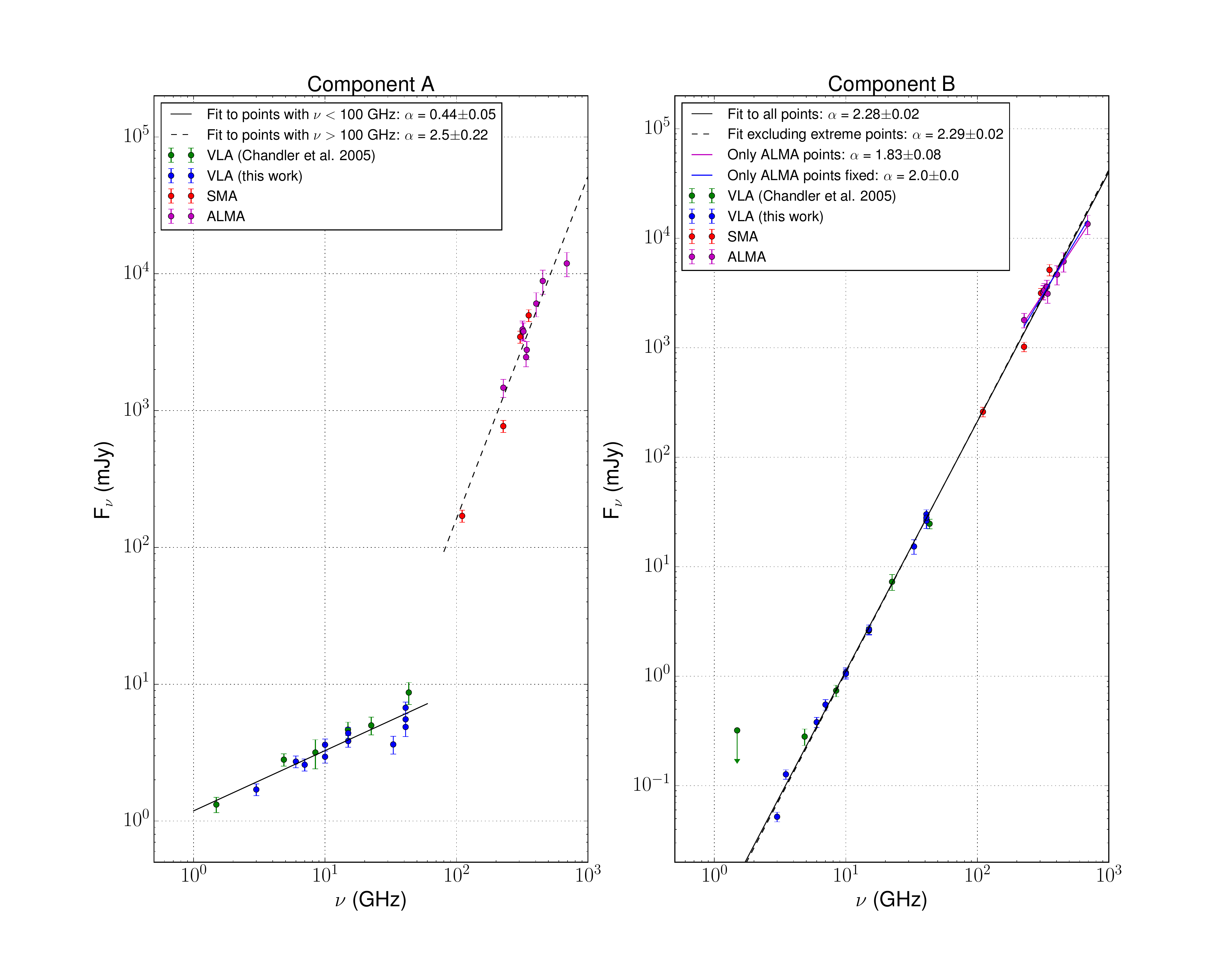}
\caption{Spectra of sources A and B derived from SMA and VLA observations (Chandler et al.\ 2005) as well as the VLA and ALMA results presented in the present paper. Fits with various power-laws, as described in the next, are shown.}
\label{fig:sed}
\end{figure*}

\section{Observations \label{sect:obs}}

\subsection{Very Large Array observations \label{sect:vla_obs}}

New Very Large Array (VLA) observations of \iras\ were obtained between 2011 and 2015 in the most extended configurations (A, BnA, and B) of the array (see Table \ref{tab:obs} for a summary). They covered the frequency range from 2 to 42 GHz, and resulted in angular resolutions better than \msec{0}{5} at all frequencies except at 3 GHz. We used the VLA calibration pipeline (version 1.3.1) provided by NRAO to flag and calibrate all our datasets with the CASA software package (version 4.2.2). The quasars 3C~286 and J1256$-$0547 were used for the bandpass calibration, while the source J1625$-$2527 was used for the complex gain calibration. The flux scale was set using the standard calibrator 3C~286. The uncertainty in the absolute flux calibration is estimated to be about  10\% for the observations up to 15 GHz. For the observations at higher frequencies, the estimated uncertainty is about 15\%. 

The continuum emission from this source is strong enough that self-calibration can be performed in order to correct for atmospheric fluctuations on timescales shorter than the cycle between the target and the phase calibrators. The calibrated visibilities were imaged using a weighting scheme intermediate between natural and uniform ({\sc robust} = 0 in CASA) to optimize the compromise between high angular resolution and high sensitivity. The resulting synthesized beams and r.m.s.\ noise levels are given in Table \ref{tab:obs}. To complete our study, we will also make use of the continuum VLA observations from \citet{Chandler2005} and \citet{Pech2010} carried out between 1986 and 2008. The reduction and calibration of these data are described in details in the original papers.

\begin{deluxetable}{ccc}[t]
\tablecaption{Deconvolved size of source B as a function of frequency \label{tab:size}}
\tablewidth{0pt}
\tablehead{
\colhead{Frequency} & \colhead{$\theta_{max}$} & \colhead{$\theta_{min}$} \\
\colhead{(GHz)}     & \colhead{(mas)} & \colhead{(mas)}}
\startdata
\decimals
10.0 & 122.0$\pm$5.9 & 42.5$\pm$39.0\\
15.0 & 137.7$\pm$5.6 & 130.0$\pm$2.5\\
22.46 & 160.6$\pm$3.2 & 141.4$\pm$5.4\\
33.02 & 192.2$\pm$6.7 & 159.9$\pm$4.5\\
41.1 & 198.6$\pm$8.2 & 172.6$\pm$6.8\\
227.0 & 416.0$\pm$28.0 & 369.0$\pm$18.0\\
318.0 & 410.0$\pm$16.0 & 376.0$\pm$15.0\\
318.0 & 407.0$\pm$13.0 & 373.0$\pm$11.0\\
323.0 & 415.0$\pm$18.0 & 380.0$\pm$18.0\\
338.2 & 374.0$\pm$13.0 & 355.0$\pm$13.0\\
342.3 & 391.0$\pm$12.0 & 365.0$\pm$12.0\\
404.0 & 410.0$\pm$16.0 & 386.0$\pm$16.0\\
453.0 & 402.0$\pm$12.0 & 370.0$\pm$10.0\\
695.88 & 408.9$\pm$9.4 & 368.1$\pm$6.9\\
\enddata
\label{tab:size}
\end{deluxetable}

\begin{figure*}[!t]
\centering
\setlength\tabcolsep{0.001pt}
\includegraphics[scale=0.45]{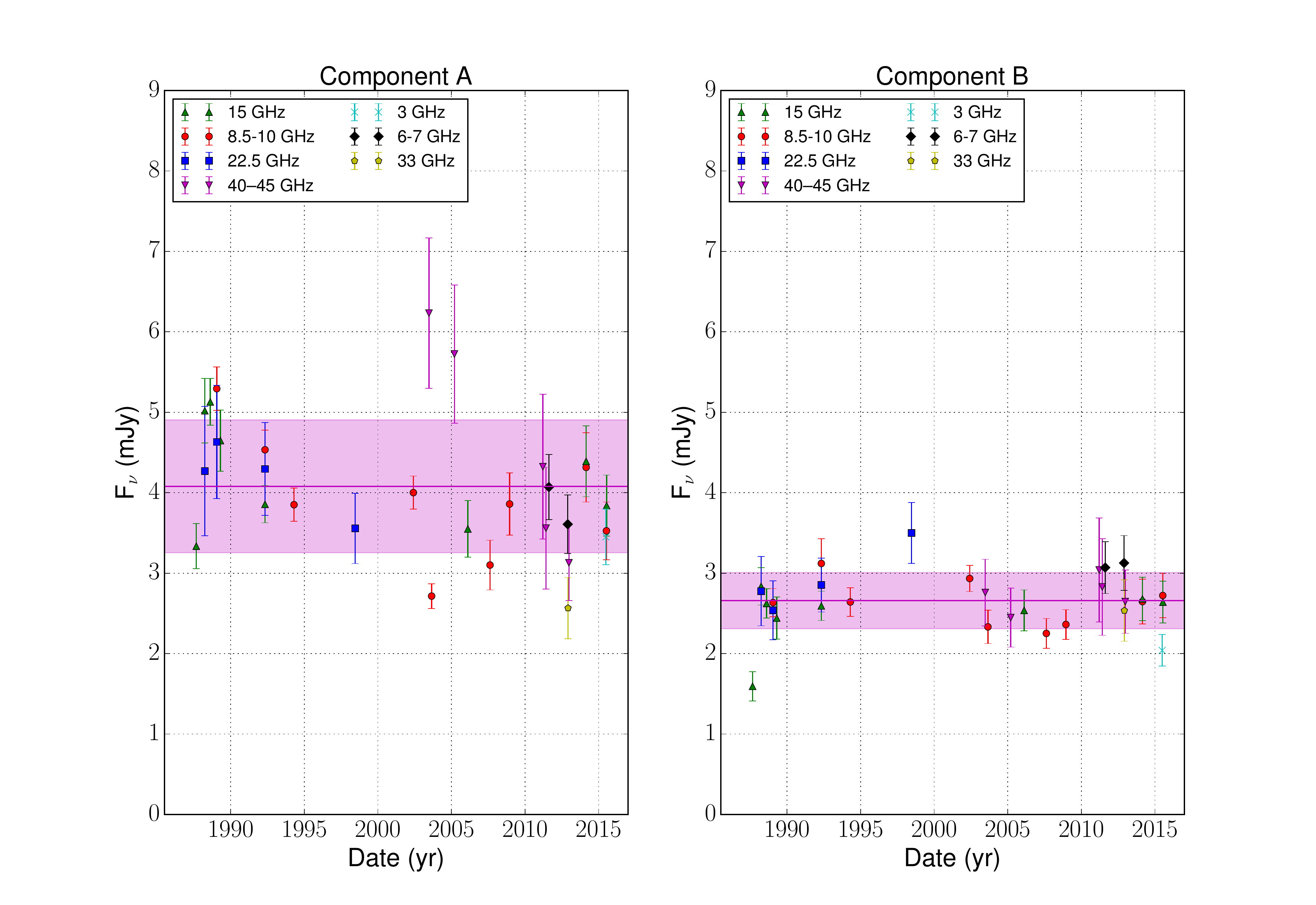}
\caption{Variability of sources A and B. For clarity, the fluxes have been renormalized to the reference frequency of 15 GHz, using the spectral indices calculated on Section \ref{sect:sed}. Different frequencies are shown with different symbols and colours. The magenta line and magenta band in each panel show the mean flux and 1-sigma dispersion around that mean, respectively.}
\label{fig:var}
\end{figure*}

\subsection{ALMA observations \label{sect:alma_obs}}

We have also used interferometric ALMA observations covering the frequency range from 230 GHz to 700 GHz (see Table \ref{tab:obs}). We selected only observations with an angular resolution better than \msec{0}{4} so that the source size of source B could be properly deconvolved (see Section \ref{sect:size} below). Images at frequencies between 227 and 453 GHz were downloaded from the database hosted by the Japanese ALMA Observatory (JAO; {\sf jvo.nao.ac.jp/portal/alma/archive.do}). These images correspond to data that have not been self-calibrated, and we assume a 10\% error on the source sizes derived from them. The absolute flux uncertainty, on the other hand, is assumed to be 15\% for frequencies below 400 GHz, and 20\% above 400 GHz. In addition, we used the Band 9 Science Verification observation of \iras\ as available on {\sf almascience.nrao.edu}, these data were self-calibrated.

\section{Results \label{sect:results}}

In Figure \ref{fig:all_images} we show all the continuum images of \iras\ obtained with the VLA between 2011 and 2015. Similar figures for observations prior to 2011 are included in \citet{Chandler2005} and \citet{Pech2010}. For the 41 GHz observations obtained during 2011, we have averaged the data taken in the B configuration on 15 and 19 March (mean epoch 2011.21) to produce the first panel of Figure \ref{fig:all_images}, and the data taken in the BnA configuration on 5 and 8 June (mean epoch 2011.42) to produce the second panel. This combining procedure results in images with higher signal-to-noise and better fidelity. To exhibit more clearly the structure of source A, labelled zooms of the images at 33 GHz (epoch 2012.92) and 15 GHz (epoch 2014.15) are provided in Figure \ref{fig:zoom}.

As expected, source B to the north-west of the system remains largely featureless in all images, and --due to its positive spectral index-- it appears much more prominent at the highest frequencies. Source A, on the other hand, appears highly structured, containing a number of sources that depends on time and frequency. These different sources correspond, on the one hand, to the A1 and A2 objects initially identified by \citet{Wootten1989} and subsequently monitored by \citet{Loinard2002}, \citet{Chandler2005} and \citet{Pech2010}, and, on the other, to the ejecta recently expelled by source A2 \citep{Loinard2007,Pech2010,Loinard2013}. 

In the following sub-sections, we will use these VLA images as well as the ALMA images described in Section \ref{sect:alma_obs} to characterize, separately, the spectra\footnote{Plots of flux density as a function of frequency, such as those that will be presented here, as often called spectral energy distributions (SED). Strictly speaking, however, an SED is a plot of {\em energy} vs.\ frequency, while plots of flux density vs.\ frequency are spectra. In the present paper, we will adhere to this definition.} of components A and B as well as their temporal variability. We will also use them to examine the size of several of the sources, and monitor the astrometry (both absolute and relative) of the different sources in the system.

\begin{deluxetable}{llll}[!t]
\tablecaption{Absolute proper motions for all sources in \iras. \label{tab:pm}}
\tablewidth{7cm}
\tablehead{
\colhead{Source} & \colhead{$\mu_{\alpha} \cos \delta$}& \colhead{$\mu_{\delta}$} \\
\colhead {} & \colhead{(mas yr$^{-1}$)}  &\colhead{(mas yr$^{-1}$)}}
\startdata
\decimals
A1         &  $-$3.0$\pm$0.4 & $-$27.9$\pm$0.8 \\
A2         &  $-$7.8$\pm$0.6 & $-$21.8$\pm$0.5 \\
A2$\alpha$ & $-$63.8$\pm$5.9 & $-$51.5$\pm$2.6 \\
A2$\beta$  & $+$53.2$\pm$3.4 & $-$17.9$\pm$1.3 \\
B          &  $-$5.7$\pm$1.4 & $-$21.0$\pm$1.1 \\
\enddata
\tablecomments{\scriptsize These values were obtained by fitting a linear and uniform proper motion to the measured absolute positions. }
\end{deluxetable}

\subsection{Spectra \label{sect:sed}}

The integrated flux densities obtained for sources A and B from resolved SMA and VLA observations between 5 and 300 GHz were previously reported by \citet{Chandler2005}. Here we extend these observations both to higher frequencies (up to 700 GHz) using ALMA observations and to lower frequencies (down to 2 GHz) using new VLA observations. We have used CASA to measure the flux densities for A and B. For source B, we used a gaussian fit since its structure at all wavelengths is well defined. On the other hand, the emission from source A is more complex, so we used the task IMFIT with multiple components in CASA to measure individual fluxes, and defined region including all sources composing A to obtain the total flux. The resulting complete spectra are shown in Figure \ref{fig:sed}.

In the case of source A, we confirm the results by \citet{Chandler2005} that the fluxes can be fitted by two separate power laws $F_{\nu} \propto \nu^{\alpha}$. At $\nu > 100$ GHz, the spectral index derived from the data is $2.50\pm 0.22$, as expected from thermal dust emission. At $\nu < 100$ GHz, the spectrum becomes shallower and is well characterized by a spectral index $\alpha=0.44 \pm 0.05$ typical of partially optically thick free-free emission from ionized jets \citep[e.g.][]{Anglada2015}. This situation is typical of deeply embedded low-mass protostars driving supersonic outflows. 

As shown in Figure \ref{fig:zoom}, source A is a composite of several components, so the spectral index derived above corresponds to the average of the spectral indices of the various sources within component A. The determination of the spectral indices of the individual sources is complicated by the blending between those sources in several of the images, and the fact that some features (particularly the ejecta) are only detected at some frequencies. To estimate the spectral indices of the individual components, we used the resolved observations obtained at 10 and 15 GHz in 2014.15. These data yield $\alpha_{A1}=0.5\pm0.2$, $\alpha_{A2}=0.7\pm0.2$, $\alpha_{A2\alpha}=-0.4\pm0.2$ and $\alpha_{A2\beta}=-0.1\pm0.2$. The positive values derived for A1 and A2 are consistent with the partially optically thick free-free emission expected for the thermal jet of young stellar objects \citep[e.g.][]{Anglada2015}. This would be consistent with the interpretation of A1/A2 in terms of a tight binary system by \citet{Loinard2007} and \citet{Pech2010}. The spectral indices of sources A2$\alpha$ and A2$\beta$, on the other hand, are consistent with the expectation ($\alpha$ = --0.1) for optically thin free-free emission, as appropriate for low density ionized ejecta. We note that \citet{Loinard2013} had found a similar value for A2$\beta$ from earlier VLA observations. It is important to point out that the ejecta contribute relatively little to the total flux of component A, so the overall spectrum of source A is dominated by the compact sources A1 and A2; this explains the overall positive index of source A.

For source B, the entire spectrum from 2 to 700 GHz can be modelled by a single power-law with spectral index  $\alpha=2.28 \pm 0.02$. Note that to constrain better the behavior of the spectrum at low frequency, we measured the flux separately in the entire 3 GHz band ($\Delta \nu$ from 2 to 4 GHz) covered by the S-band observations, and in sub-bands from 2 to 3 GHz (centered at 2.5 GHz), and from 3 to 4 GHz (centered at 3.5 GHz). Source B is only detected in the entire band and in the higher frequency (3.5 GHz) sub-band. It is not detected in the lower frequency sub-band centered at 2.5 GHz. This is fully consistent with the spectral index derived from the entire spectrum (Figure \ref{fig:sed}). To examine whether or not the highest and lowest frequency points (at 695 GHz and 2.0 GHz) suggested a departure from a single power law, we also fitted the data ignoring these two extreme points. This fit is very nearly identical to that obtained including all points. Nevertheless, the highest frequency (ALMA) data points do tend to fall below the power-law fit. Using only the ALMA points yields a somewhat smaller spectral index of 1.83$\pm$0.08 (Figure \ref{fig:sed}). The value of the spectral index of source B ($\alpha = 2.28 \pm 0.02$) is consistent with thermal dust emission, and is statistically incompatible with the maximum value ($+$2.0) for optically thick free-free emission. Instead, the spectrum suggests that the emission is due to thermal dust radiation over the entire range from 700 GHz down to 2 GHz, with no detectable free-free contribution even at the lowest frequency (further evidence for this conclusion will be presented in section \ref{sect:size}). This is, to the best of our knowledge, a unique situation. A number of other young stellar objects \citep[e.g.][]{Brogan2016} have been observed to be dominated by thermal dust emission down to $\sim$ 10 GHz, but never down to 2 GHz.

\citet{anglada2018} reviewed the properties of centimeter free-free emission from jets driven by young stellar objects. They confirmed the existence of an empirical relationship between the bolometric luminosity of the object and the intensity of the free-free emission from its thermal jet. \iras\ has a total bolometric luminosity of about 25 \Lsun\ \citep[][when scaled to the distance of 141 pc adopted here]{makiwa2014,crimier2010,correia2004,schoier2002}. The empirical relationship derived by \citet{anglada2018} for that luminosity predicts a total centimeter free-free flux density for \iras\ of order 4 mJy (to within a factor of a few). This is indeed, the total centimeter flux density measured here for the sum of source A and B. \cite{Jacobsen2018} estimate that around 85\% of that the total luminosity of \iras\ is attributable to source A, while the remaining 15\% come from source B. Thus, one would expect centimeter free-free flux densities of about 3.4 and 0.6 mJy for source A and B respectively. The measured centimeter flux density of source A is consistent with this expectation, but that of source B is not. Indeed, the predicted contribution from dust at $\sim$ 6--7 GHz (from the power law derived from the spectrum of source B) is about 0.5 mJy. Thus, the total flux expected for source B, if it fell on the empirical relationship of \citet{anglada2018}, would be about 1 mJy. In contrast, the measured flux density of source B is only about 0.5 mJy, and is fully compatible with the predicted thermal dust emission alone. The dispersion of the data points around the fit to the spectrum is of order 10\%, so a reasonable upper limit to the free-free contribution is about 0.05 mJy, or about a tenth of the expected flux for a source with that luminosity. We conclude that source B does not drive a powerful ionized outflow. 

\begin{figure}[!t]
\centering
\setlength\tabcolsep{0.001pt}
\includegraphics[scale=0.47]{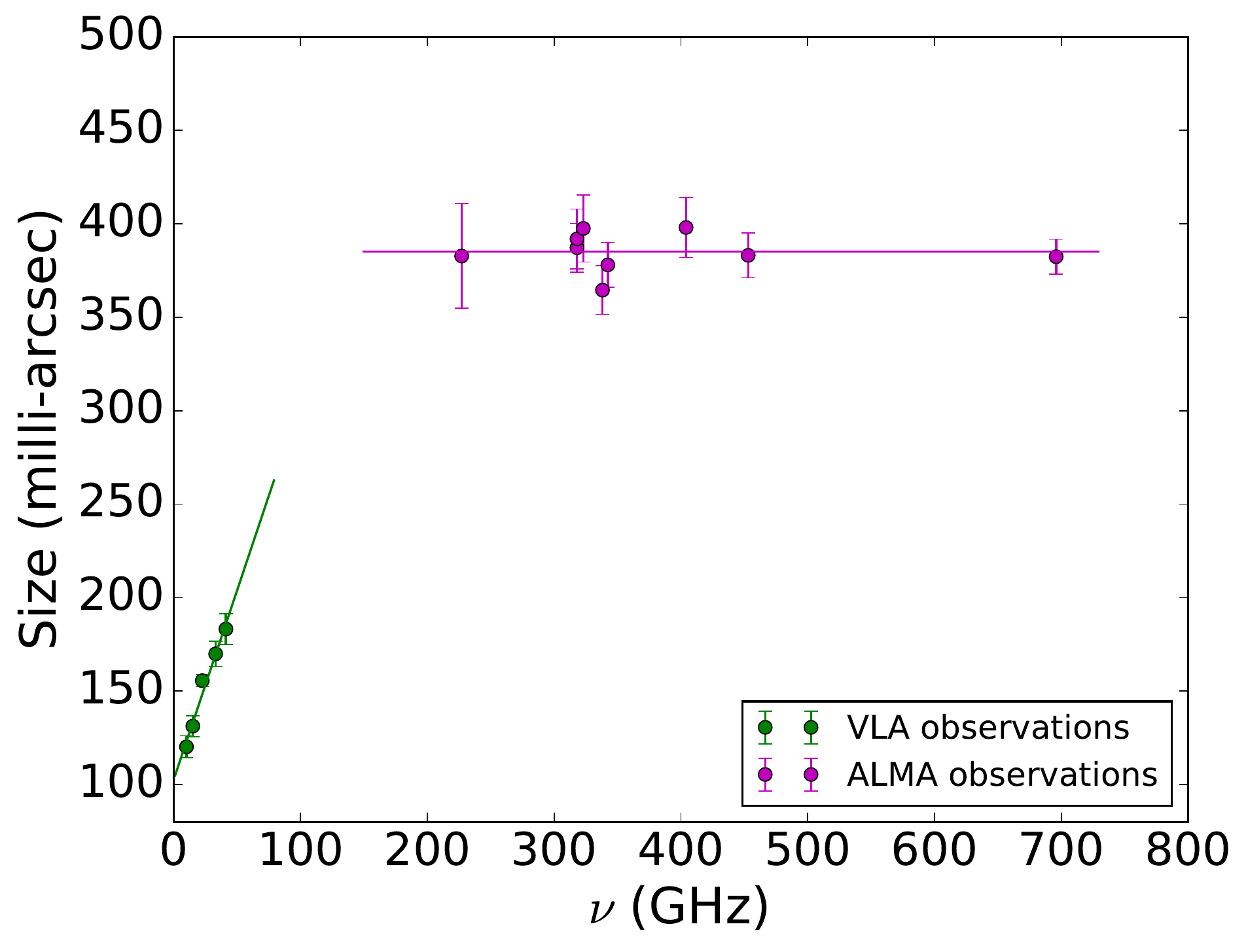}
\caption{Size of source B as a function of frequency. The size increases linearly up to 50 GHz, while it remains roughly constant at frequencies higher than 200 GHz.}
\label{fig:size}
\end{figure}

\subsection{Variability \label{sect:var}}

\citet{Chandler2005} have shown that the centimeter radio flux of source A is somewhat time variable, with a variability of order $\sim$ 50\% at any given frequency. This result is confirmed by the new VLA observations. To characterize the radio variability of sources A and B separately, we considered the fluxes measured with the VLA at all available epochs, and we re-scaled them to a single reference frequency (chosen, arbitrarily, to be 15 GHz) using the spectral indices of sources A and B measured in Section \ref{sect:sed}. Specifically, we multiplied the fluxes measured at a given frequency $\nu$ by $(\textrm{15 GHz} / \nu)^{\alpha}$ where $\alpha$ = 0.45 for source A and $\alpha$ = 2.28 for source B. This eliminates the variations due to frequency and reveals intrinsic variability more explicitly. The results are shown in Figure \ref{fig:var} where it is immediately obvious that source A is more variable than source B. Quantitatively, the mean flux of source A (expressed at the reference frequency of 15 GHz) is found to be 4.08 mJy with a dispersion of 0.83 mJy around that mean. In comparison, the mean flux of source B is 2.66 mJy with a dispersion of only 0.35 mJy. Indeed, with the exception of the very first data point (at 15 GHz in 1987), the statistical distribution of all measured fluxes are compatible with the mean value. As discussed by \citet{Chandler2005}, it is quite possible that the first data point is affected by an absolute flux calibration issue.

The variability of source A is certainly expected given that the centimeter emission traces mass ejection which, as evidenced by the recent ejecta A2$\alpha$ and A2$\beta$, is clearly time dependent. The flux of source A2$\alpha$ at 10 GHz, for instance, is found to have faded by 30 to 40\% between 2014.15 and 2015.53.

\subsection{Source sizes \label{sect:size}}

The components A1 and A2 of source A remain unresolved in our observations, and the ejecta source A2$\beta$ is too heavily blended with A1 for a reliable size to be measured. The ejecta A2$\alpha$, on the other hand, is well separated from the other components of source A in the 2014.15 and 2015.53 observations at 10 and 15 GHz. Epoch 2014.15 is especially interesting in this respect, because source A2$\alpha$ is detected with high signal to noise ratio. In comparison, the 2015.53 detections have a significantly lower signal to noise ratio, because these observations are intrinsically shallower, and because A2$\alpha$ has faded significantly between 2014.15 and 2015.53 (see Section \ref{sect:var}). The deconvolved size of source A2$\alpha$ in 2014.15 is found to be (\msec{0}{44}$\pm$\msec{0}{04}) $\times$ (\msec{0}{11}$\pm$\msec{0}{05}) at position angle (116$^\circ \pm$5$^\circ$) at 15 GHz, and (\msec{0}{47}$\pm$\msec{0}{04}) $\times$ (\msec{0}{17}$\pm$\msec{0}{06}) at position angle (113$^\circ \pm$7$^\circ$) at 10 GHz. Thus, we find no statistical evidence for a different size at the two frequencies, as expected for optically thin free-free emission. On the other hand, source A2$\alpha$ is found to be significantly larger than in the 2007 and 2008 observations reported by \citet{Pech2010} who quote (\msec{0}{21}$\pm$\msec{0}{06}) $\times$ (\msec{0}{08}$\pm$\msec{0}{04}) at 8.5 GHz for epoch 2008.95. The position angle was poorly constrained for that epoch. 

Source B is well resolved at all frequencies above 10 GHz in the interferometric observations presented here (Table \ref{tab:size}). As noticed by \citet{Chandler2005}, the size increases with frequency in the centimeter regime, from about 120 mas at 10 GHz to about 185 mas at 40 GHz (Figure \ref{fig:size}). This tendency must continue between 40 and 230 GHz (at least in part of that frequency range), since the size measured at the latter frequency is 380 mas. As \citet{Chandler2005} pointed out, this increase of the size with frequency further confirms that the emission is dominated by partially optically thick thermal dust emission. Once in the millimeter regime, however, the size remains constant at 385$\pm$5 mas (54.3$\pm$0.7 AU assuming a distance of 141 pc -- \citealt{Dzib2018}), suggesting that the source is entirely optically thick. This is consistent with the spectral index, equal to 2.0 within 2$\sigma$, observed at ALMA frequencies (Figure \ref{fig:sed}). At lower frequencies, the emission is a mixture of an optically thick core (with spectral index 2.0) and a optically thin outer region (where the spectral index is higher than two). This combination naturally results in a spectral index larger than 2.0. In this scheme, the size measured at millimeter wavelengths corresponds to the full extent of source B.

\begin{figure*}
\begin{tabular}{l l l}
\includegraphics[width=0.33\textwidth,trim= 0 0 0 0,clip]{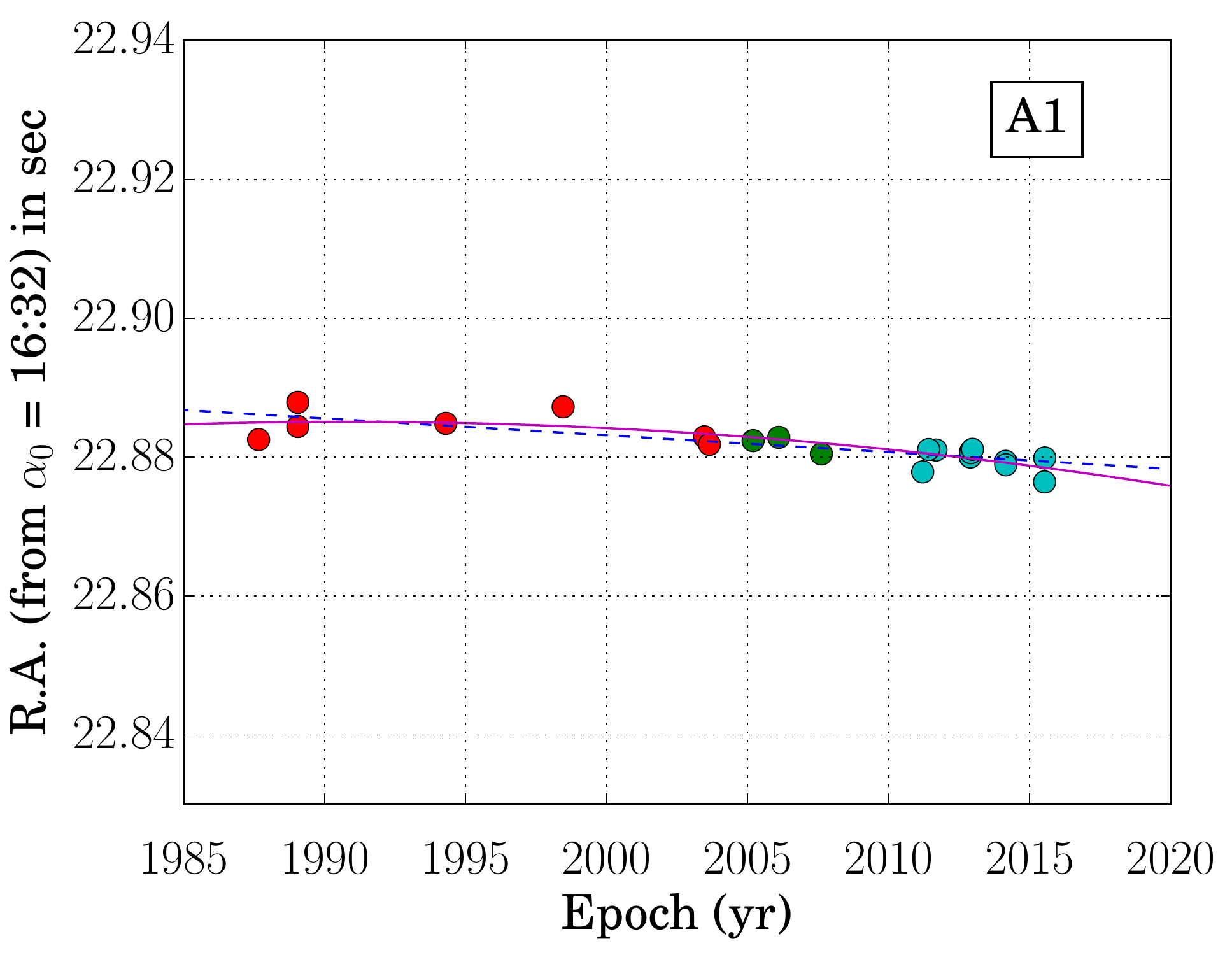} & 
\includegraphics[width=0.33\textwidth,trim= 0 0 0 0,clip]{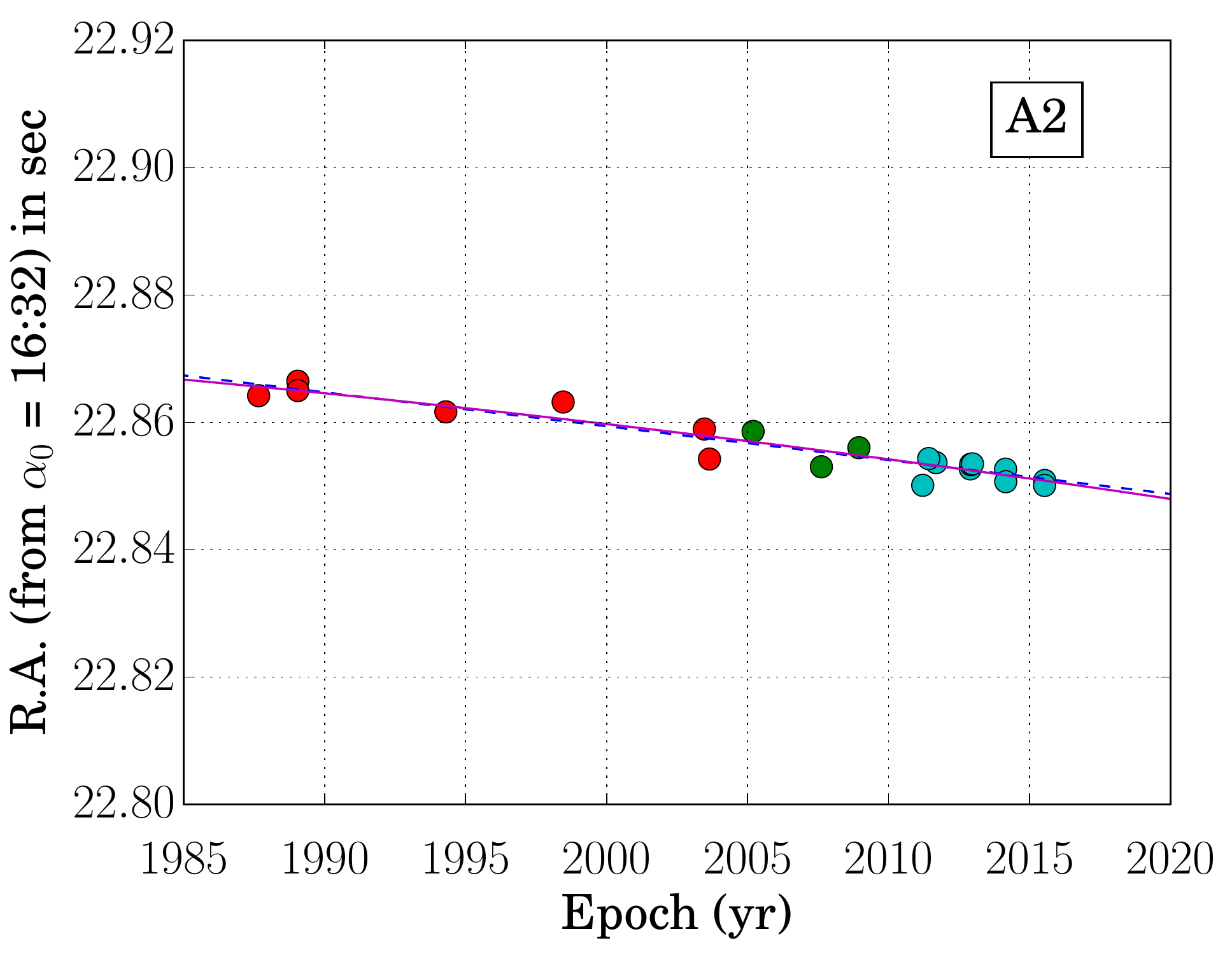} &
\includegraphics[width=0.33\textwidth,trim= 0 0 0 0,clip]{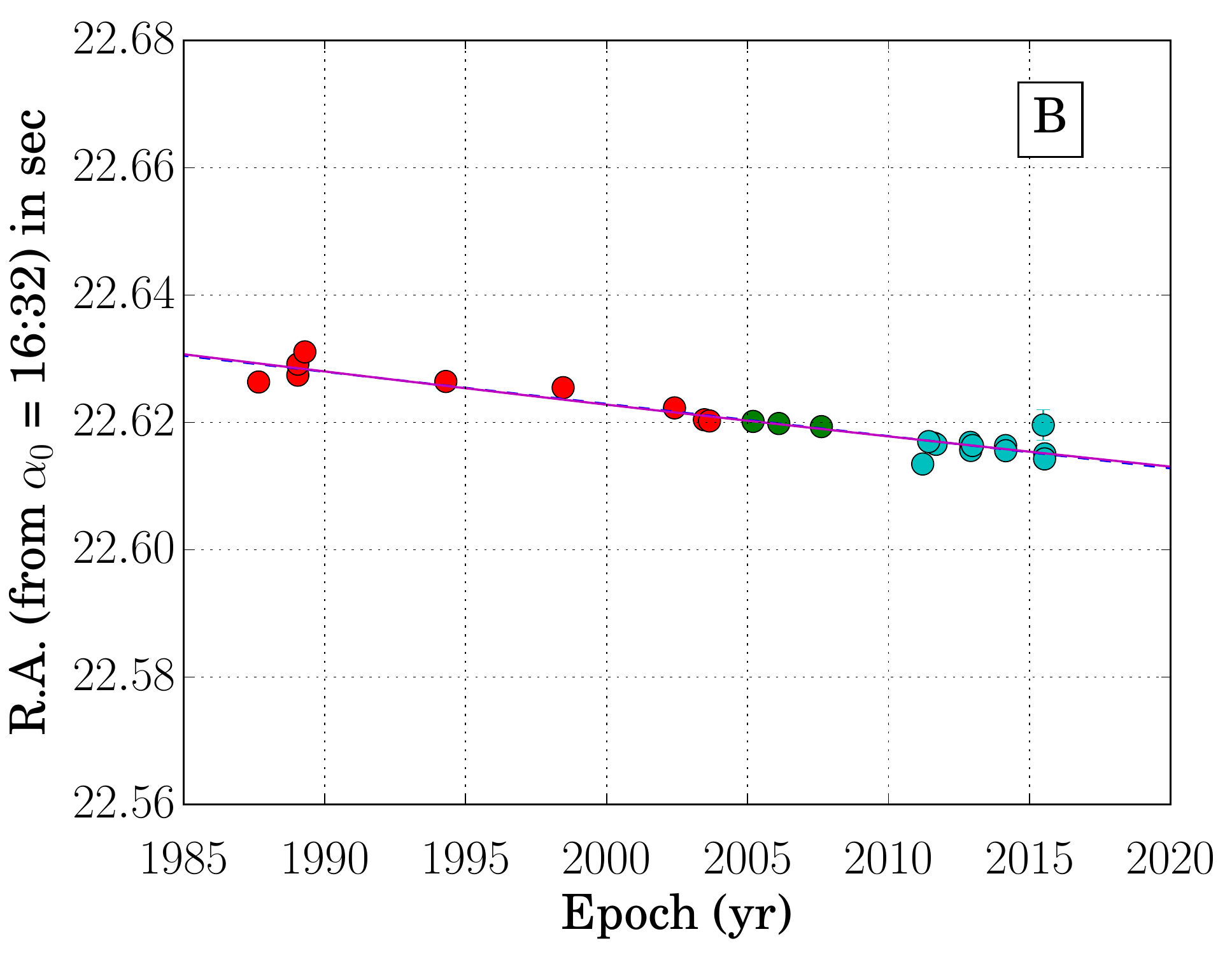}  \\
\includegraphics[width=0.33\textwidth,trim= 0 0 0 0,clip]{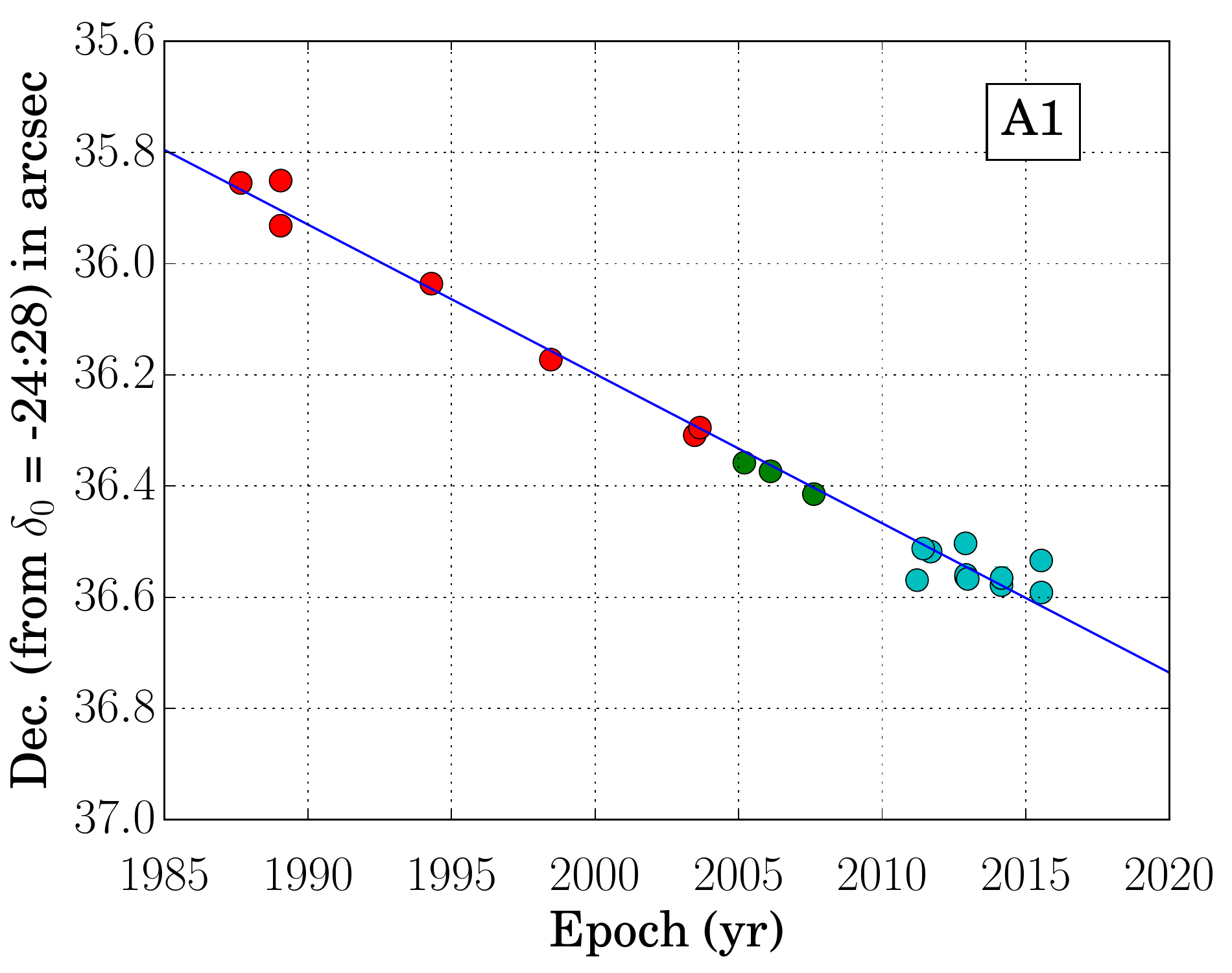} &
\includegraphics[width=0.33\textwidth,trim= 0 0 0 0,clip]{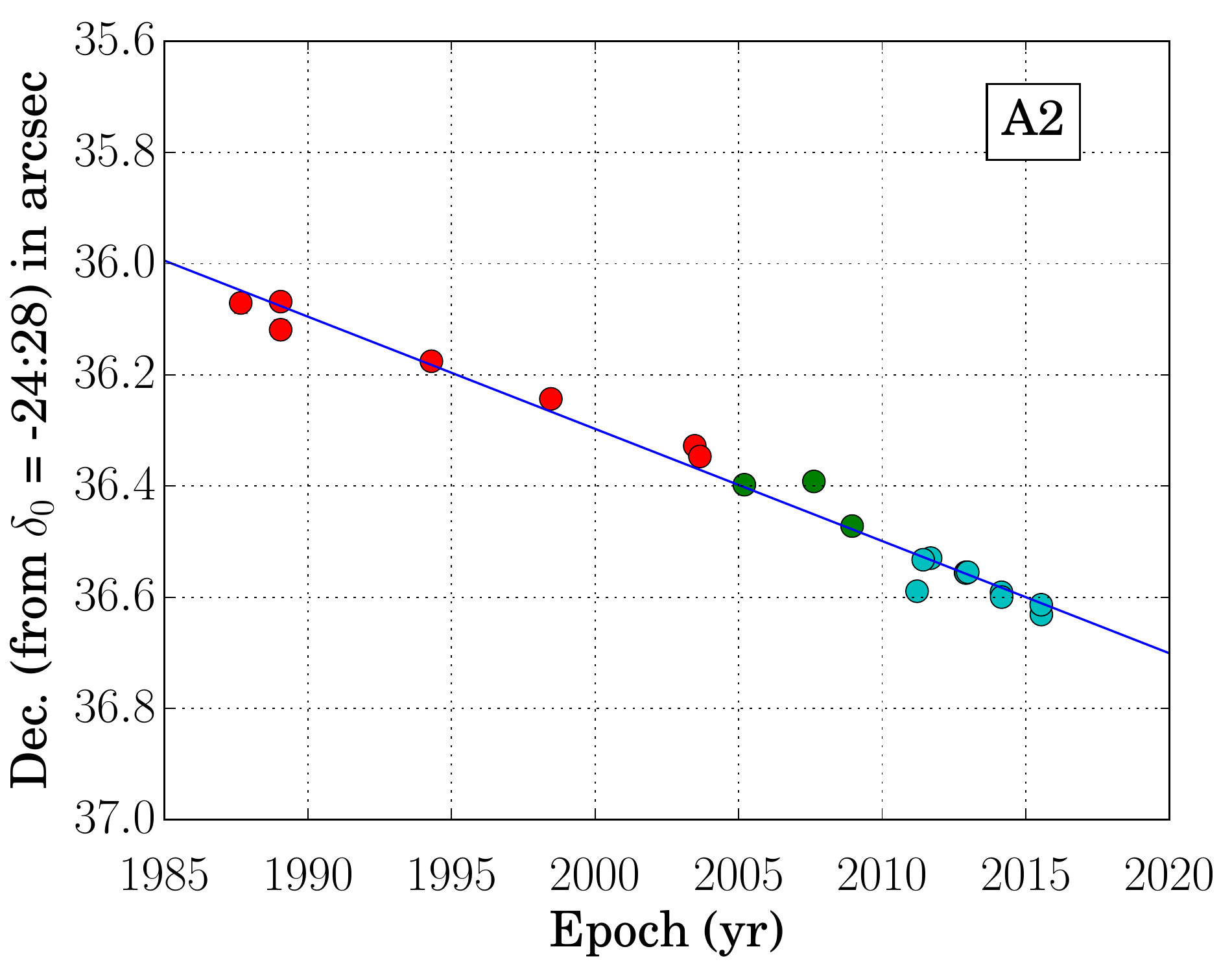} &
\includegraphics[width=0.33\textwidth,trim= 0 0 0 0,clip]{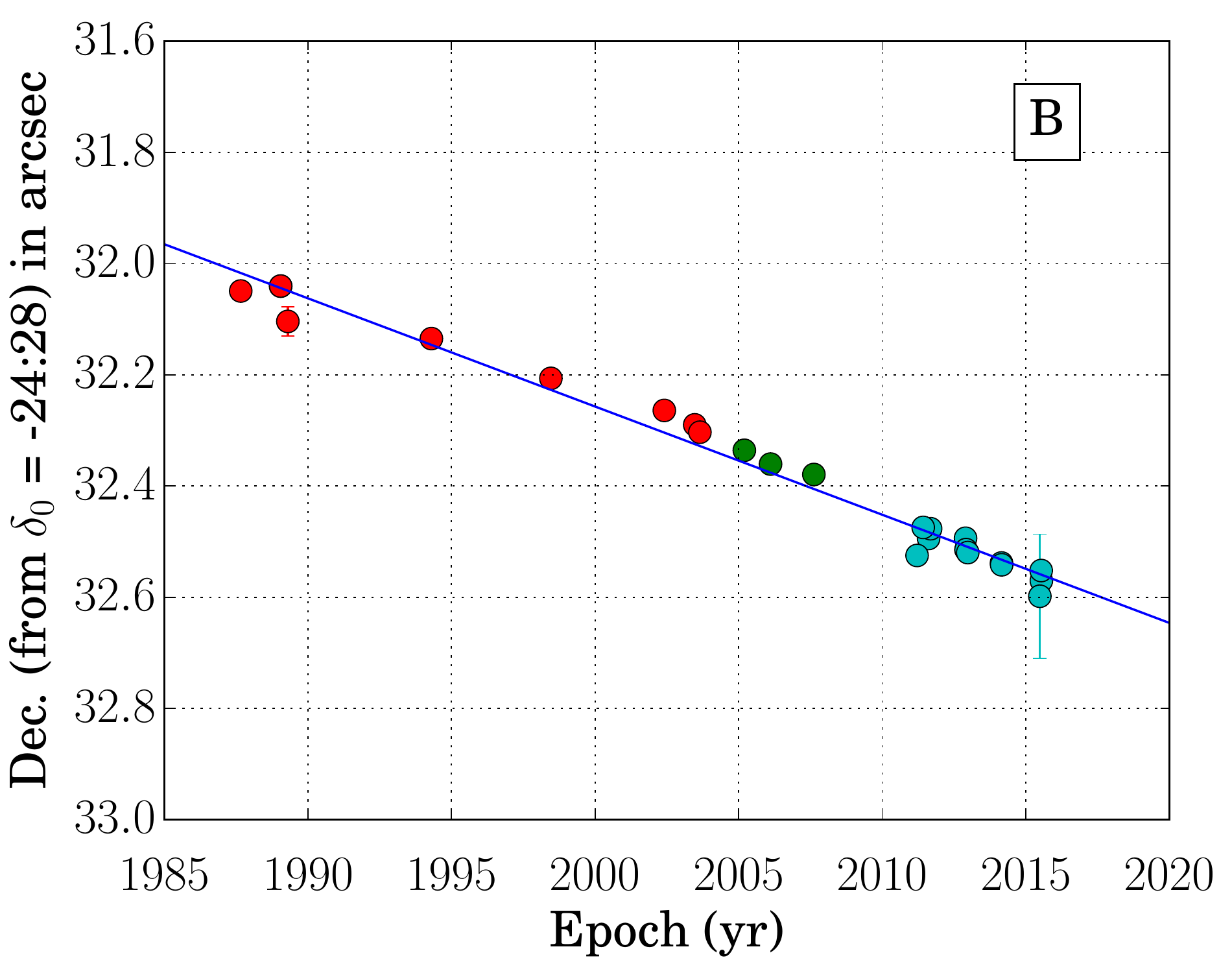}  \\
\end{tabular}
\caption{Absolute position as a function of time for sources A1, A2 and B (from left to right). The red points correspond to the observations from Chandler et al.\ (2005), the green points represent the observations from Pech et al.\ (2010) and the cyan points correspond to this work. The blue lines show the best fits with a first order polynomial. \label{fig:pm_abs}}
\end{figure*}

\subsection{Astrometry}

The relative motion between sources A1 and A2 was first reported by \citet{Loinard2002} from VLA observations at 8.4 GHz obtained between 1989 and 1994. \citet{Chandler2005} and \citet{Pech2010} expanded that study to include source B, using data obtained between 1987 and 2009. \citet{Pech2010} also monitored the motion of the ejecta A2$\alpha$ and A2$\beta$ that were first reported by \citet{Loinard2007}.

\begin{figure*}
\centering
\begin{tabular}{c c}
\includegraphics[width=0.4\textwidth,trim= 0 0 0 0,clip]{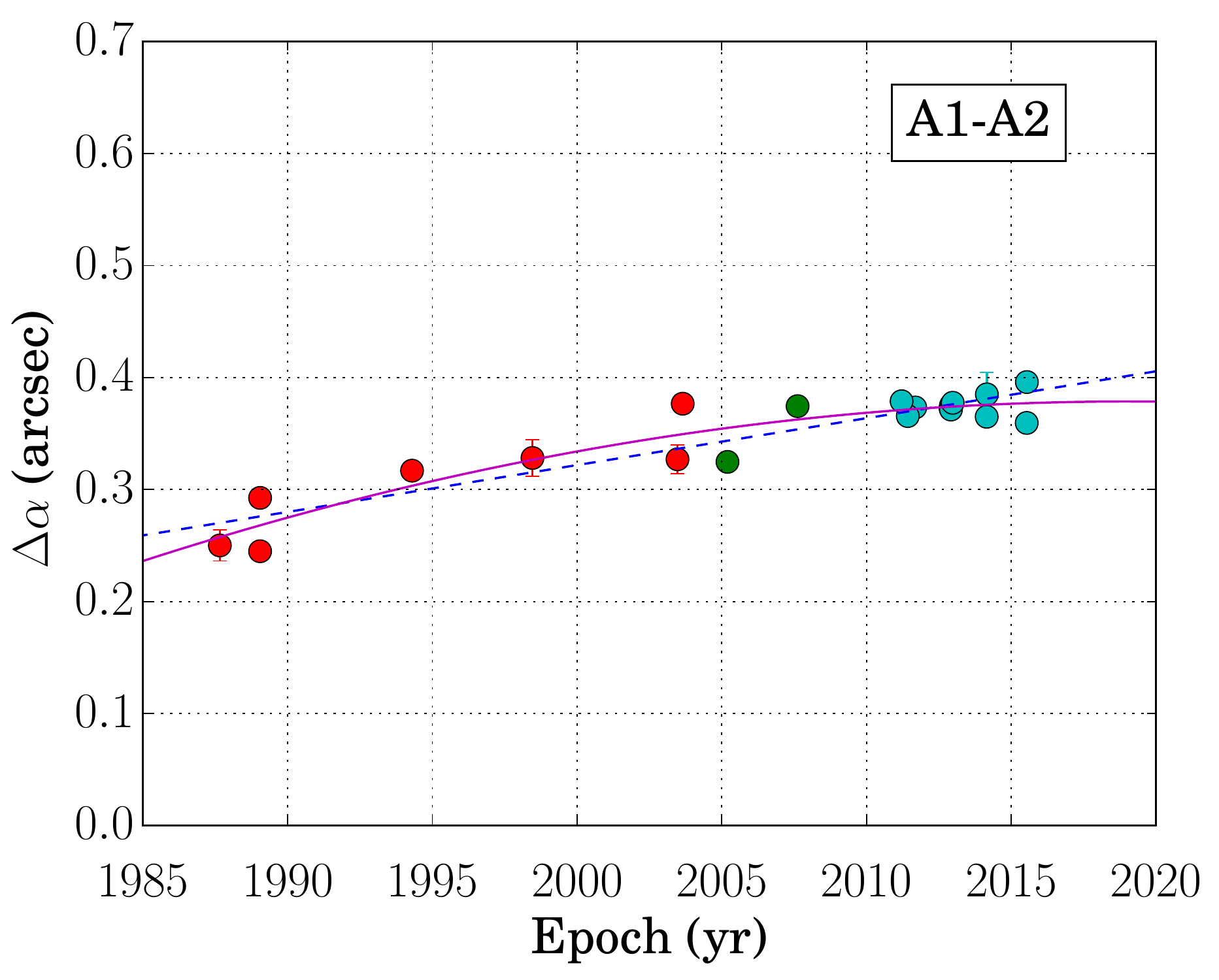} &
\includegraphics[width=0.41\textwidth,trim= 0 0 0 0,clip]{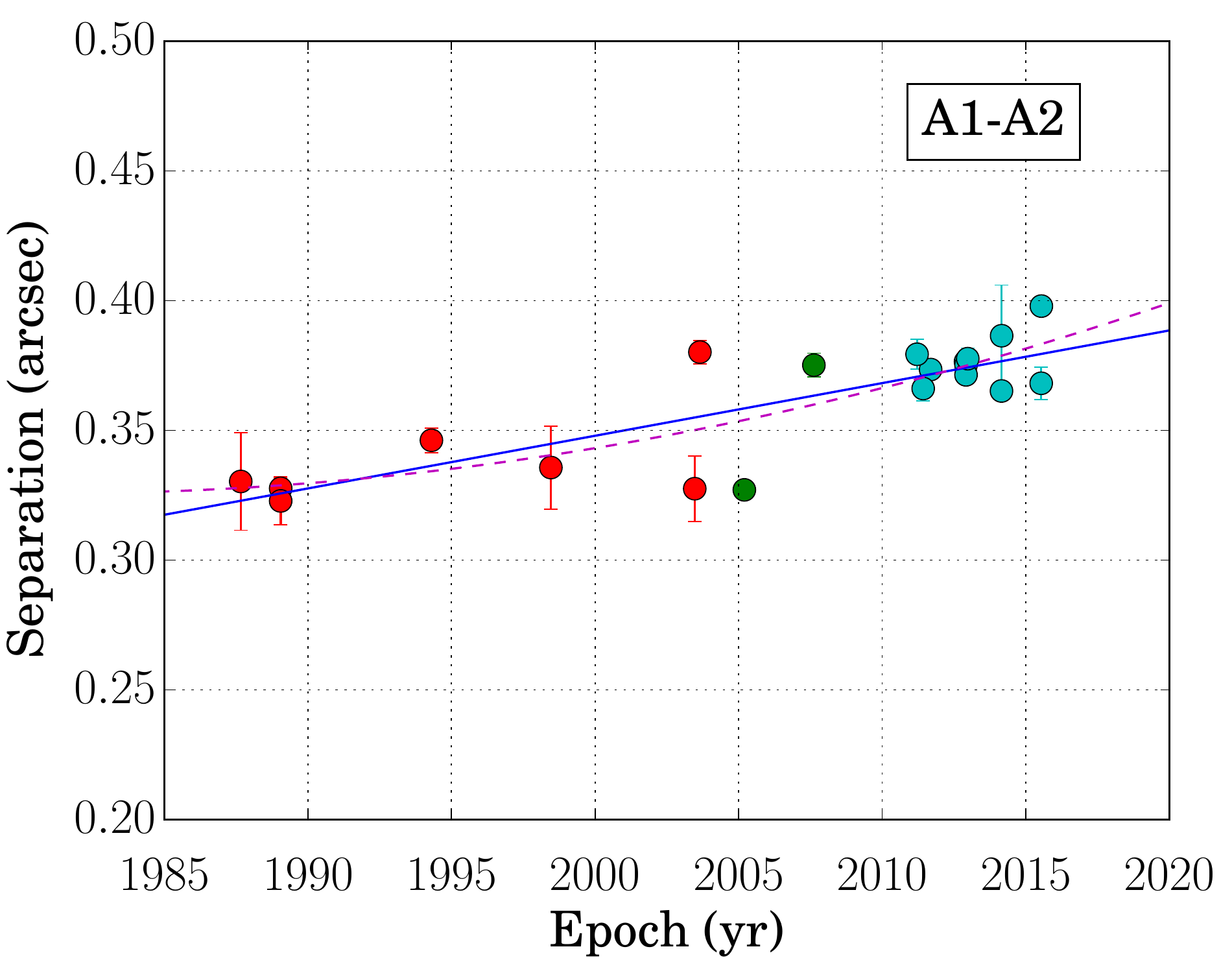} \\ 
\includegraphics[width=0.41\textwidth,trim= 0 0 0 0,clip]{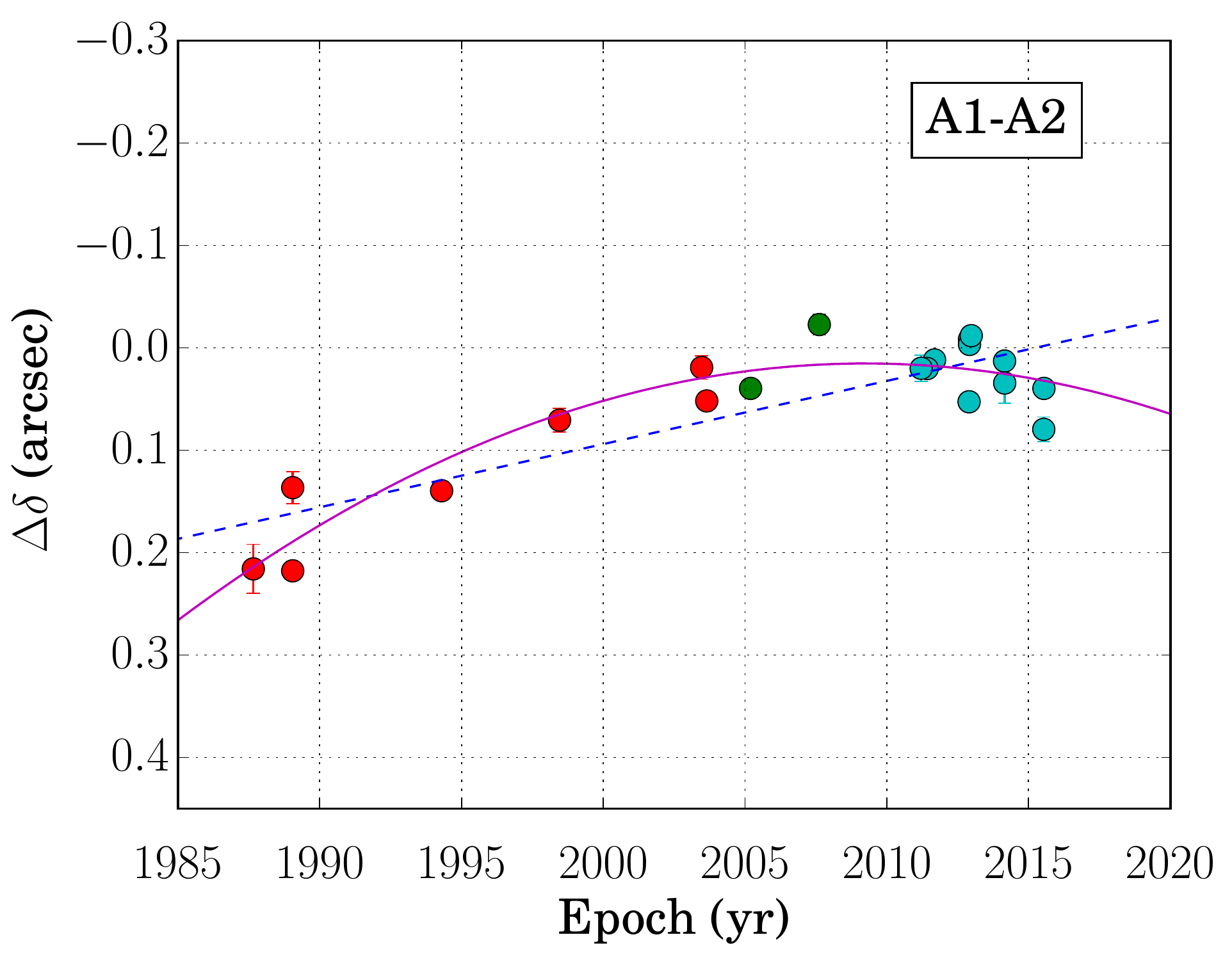} &
\includegraphics[width=0.4\textwidth,trim= 0 0 0 0,clip]{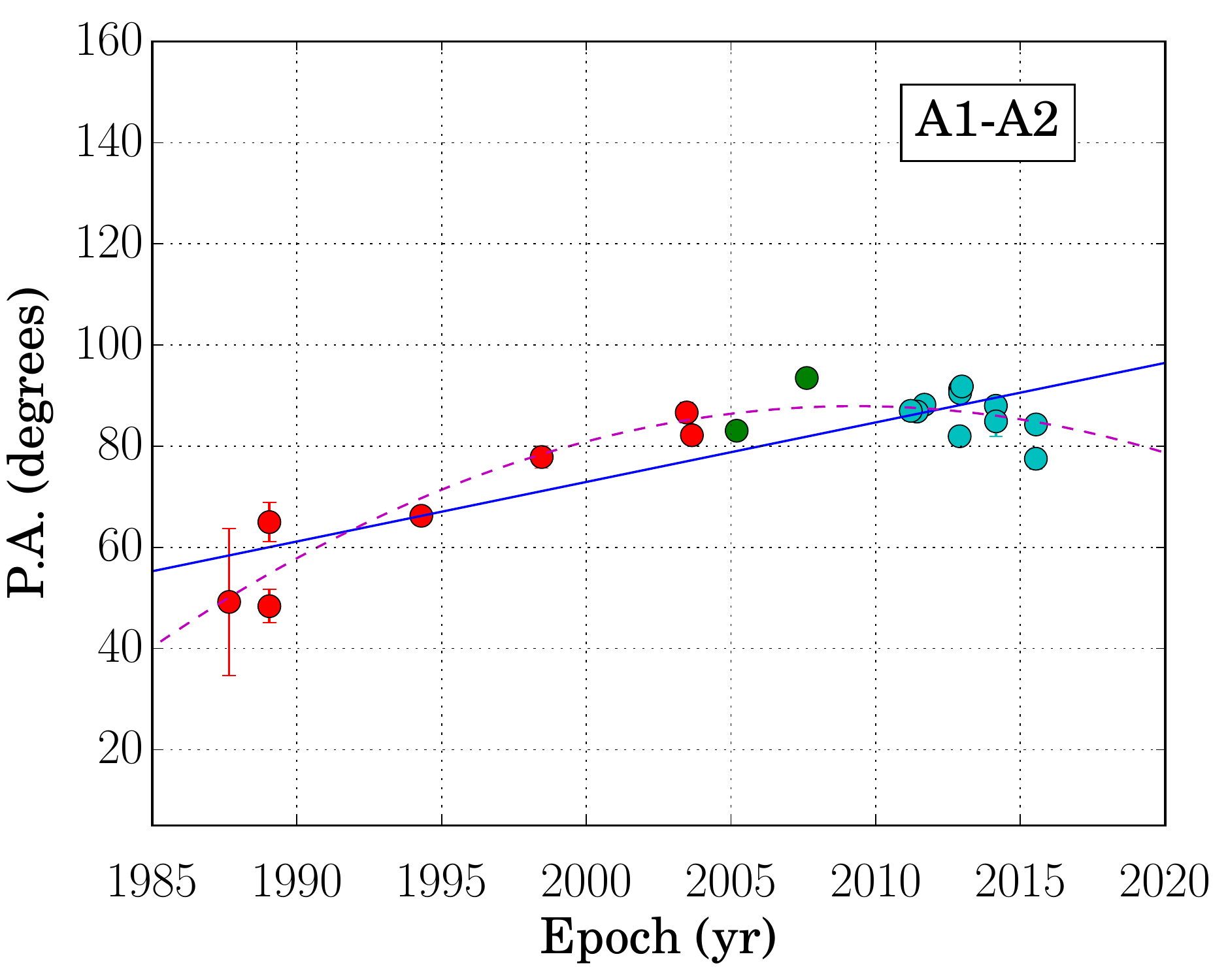}  \\
\end{tabular}
\caption{Relative motion between A1 and A2 in right ascension (upper left panel), declination (lower left panel), separation (upper right panel) and position angle (lower right panel). The blue lines (solid and dashed) show first order polynomial fits, while the magenta lines show a second order fit. The symbol colours have the same meaning as in Figure \ref{fig:pm_abs}. \label{fig:A1A2}}
\end{figure*}

The absolute position as a function of time for sources A1, A2, and B are shown in Figure \ref{fig:pm_abs}, and the derived proper motions are given in Table \ref{tab:pm}. While the positions are reasonably well fitted by linear and uniform proper motions, we find some evidence for a deviation from this expected behaviour in the data. Indeed, we find that a second order polynomial does reproduce the data significantly better. However, since accelerated absolute proper motions are unphysical in our case, we will use the linear fits to characterize the proper motions of the sources in \iras.
These fits confirm that sources A2 and B exhibit very similar proper motions. Their mean value ($\mu_\alpha \cos \delta$ = $-$ 7.5$\pm$0.8 mas yr$^{-1}$; $\mu_\delta$ = $-$21.7$\pm$0.7 mas yr$^{-1}$) provides a good proxy for the relative motion between the Sun and \iras. We note that this motion is largely dominated by the reflex Solar motion. Assuming the values provided by \citet{schonrich2010} for the Solar motion, we calculated that the reflex Solar motion for the direction of \iras\ is $\mu_\alpha \cos \delta$ = --8.1. mas yr$^{-1}$, $\mu_\delta$ = --19.4 mas yr$^{-1}$. In other words, the motion of \iras\ relative to the LSR is 0.6 $\pm$ 0.8 mas yr$^{-1}$ and 2.3 $\pm$ 0.7 mas yr$^{-1}$ in right ascension and declination, respectively.\\ 
The proper motion of source A1, on the other hand, is somewhat different. To examine the relative motion between A1 and A2, Figure \ref{fig:A1A2} shows the right ascension and declination offsets between A1 and A2 as a function of time (left column), as well as their separation and relative position angle as a function of time (right column). This shows that the separation between A1 and A2 is slowly increasing, from about \msec{0}{33} ($\equiv$ 46.5 AU) in the late 1980s to about \msec{0}{38} ($\equiv$ 53.6 AU) in 2015. Somewhat unexpectedly, the relative position angle between A1 and A2, which has increased by about 40$^\circ$ between the late 1980s and the early 2000s, has now started to decrease. We will discuss possible explanations in Section \ref{sect:disc}.

The relative motion between the ejecta A2$\alpha$ and A2$\beta$ and their driving source A2 are, respectively 63$\pm$3 mas yr$^{-1}$ at a position angle $+$62$\pm$10$^\circ$, and 62$\pm$3 mas yr$^{-1}$ at a position angle $+$266$\pm$10$^\circ$. Both position angles are in agreement with the orientation (P.A.\ $\sim$ 65$^\circ$) of the NE-SW outflow driven from within component A \citep[e.g.][]{Mizuno1990}. We conclude, as did \citet{Loinard2007} and \citet{Pech2010}, that this specific outflow is driven by A2. The amplitudes of the proper motions of A2$\alpha$ and A2$\beta$ relative to A2 are very similar, and imply transverse velocities of order 45 km s$^{-1}$. 

\section{Discussion \label{sect:disc}}

Starting from their radio and millimeter properties described in section \ref{sect:results}, we now proceed to discuss the nature of the various sources in \iras. 

\subsection{The protostellar source A2}

Source A2 is clearly associated with a protostar. Its centimeter emission is compact and exhibits a positive spectral index ($\alpha$ = $+$0.7$\pm$0.2) typical of the thermal jets driven by low-mass protostars \citep{Anglada2015}. Moreover, it has recently ejected two condensations (A2$\alpha$ and A2$\beta$) that are observed to move symmetrically away from A2 along a position angle that corresponds to that of the NE--SW outflow first observed by \citet{Mizuno1990}. Incidentally, we note that this association between A2 and the NE--SW outflow is the only one that can currently be ascertained. The other outflows known to exist in the system cannot yet be unambiguously associated with specific protostars. 


\subsection{The ejecta A2$\alpha$ and A2$\beta$}

The sources A2$\alpha$ and A2$\beta$ were ejected from A2 around 2005 \citep{Pech2010}. The present observations confirm that the emission mechanism of the centimeter emission is optically thin thermal bremsstrahlung. Furthermore, the size of the well resolved source A2$\alpha$ is found to have increased significantly between 2009 and 2014, demonstrating that it is diluting into its surroundings. It is, to our knowledge, the first time that such an expansion of a protostellar ejecta is observed.  


\subsection{Source A1}

The nature of source A1 has been a matter of some debate. \citet{Chandler2005} proposed that A1 is a {\em shock feature} that corresponds to the impact on the surrounding medium of a jet driven by a protostar located close to A2. Initially, it was thought that A2 itself might be the source of that jet, but the recent ejections of A2$\alpha$ and A2$\beta$ by A2 demonstrate that the jet from A2 does not point in the direction of A1. Thus, a different protostar (presumably a very close companion of A2) would have to be at the origin of the jet. \citet{Chandler2005} interpreted the relative motion between A1 and A2 as a consequence of the precession of that putative jet, and attribute the strong precession to the binary nature of the driving source. This interpretation would explain naturally the recent reversal in the behaviour of the relative position angle between A1 and A2 (Figure \ref{fig:A1A2}d). On the other hand, this interpretation is difficult to reconcile with the permanency of source A1 as a compact source of nearly constant flux over the last three decades. Indeed, A1 has been observed to move (on the plane of the sky) by about 35 AU over the course of the last 30 years. If A1 is interpreted as a shock feature from a precessing jet, then the properties of the shocked ambient material would be expected to change drastically over that time span, and this could easily have resulted in significant changes in the radio properties of A1. 

The alternative interpretation, that has been put forward by \citet{Loinard2007} and \citet{Pech2010}, is that A1 is a protostar which, together with A2, form a tight binary system. This interpretation explains more naturally the steadiness of A1 over the last 30 years. In addition, the spectral index of A1 ($\alpha$ = $+$0.5$\pm$0.2) is consistent with that of a protostar driving a thermal jet \citep{Anglada2015}. On the other hand, the recent behaviour of the relative position angle between A1 and A2 is clearly incompatible with a Keplerian orbit. 

Here, we propose a third possibility, closely related to the tight binary interpretation just mentioned: that A1 is a member of a very tight binary system which, together with A2, form a hierarchical triple system. In that scheme, the relative motion between A1 and A2 would be the combination of the motion of the center of mass of the tight binary around A2, and the motion of A1 around the center of mass of that putative very tight binary. This would result in ``epicycles'' that could easily explain the complex behavior of the relative position between A1 and A2. Future astrometric observations will be needed to test this possibility, but we note that the existence of three protostars in source A is not unexpected. As we described in Section \ref{sect:intro}, two outflows (one oriented E--W and the other oriented NE--SW) have long been known to be powered from within A, and \citet{Girart2014} proposed that a third outflow (oriented NW--SE) also originated from source A. This would clearly require the existence of three protostars in source A.

\subsection{Source B}

Two of the radio properties of source B are quite extraordinary. The first one is that its spectrum can be described as a single power law from 2 to 700 GHz, with a spectral index (2.28$\pm$0.02) that is incompatible with thermal bremsstrahlung. There is some evidence that the spectrum becomes somewhat shallower (and compatible with 2.0) at the highest frequencies. Since the emission in the millimeter/sub-millimeter regime is clearly due to dust, the emission in the centimeter regime is likely of the same origin. This is confirmed by the second remarkable property of source B: that its size increases with frequency between 10 and 50 GHz, but stays constant from 230 to 700 GHz. We interpret this as evidence that the source becomes increasingly optically thick as the frequency increases until, in the millimeter and sub-millimeter regimes it is entirely optically thick. Given this conclusion, the brightness temperature measured at the central position of source B in the well-resolved images obtained with the VLA at 33 GHz ($T_b$ = 870 K) and 41 GHz (T$_b$ = 700 K) provide a stringent lower limit\footnote{The brightness temperature becomes equal to the kinetic temperature when (i) Local Thermodynamical Equilibrium (LTE) conditions apply (this is very likely at the high densities found near the center of source B), and (ii) when the optical depth tends to the infinite.} to the kinetic temperature in the central parts of source B.

The brightness temperature derived from the ALMA observations is somewhat lower ($\sim180$ K). For instance, from the Science Verification Band 9 ALMA data at 700 GHz with a spatial resolution of $\sim0.3''$, we have $T_b$ = 182 K in the central pixel. In principle, this could be due to beam dilution since the beam area of the ALMA data used here is typically five to ten times larger than that of the VLA observations at 33 and 41 GHz. However, we have obtained higher angular resolution maps at $\sim300$ GHz (E.\ Caux private communication), which show that the brightness temperature in the central pixel at \msec{0}{12} angular resolution is also about 185 K. This strongly suggests that beam dilution is not the cause of the lower brightness temperature measured with ALMA as compared with the VLA. Instead, a second effect could naturally explain the difference in brightness temperature. Given the dependence of opacity on wavelength, we expect lower frequency observations to probe regions deeper inside source B, and higher frequency data to only see the relatively outer layer, where the temperature is lower. 

We conclude that source B is a dusty structure with a very high central density (this is required to explain the high opacity of source B) and high central temperature (at least 1,000 K). The lack of excess centimeter emission that would indicate the presence of a strong ionized wind and the absence of high-velocity molecular emission around source B \citep{Mizuno1990,Girart2014} might indicate that source B is at a very early stage of its protostellar evolution.

\section{Conclusions and perspectives}

In this paper, we presented multi-epoch continuum observations of the Class 0 protostellar system \iras\ taken with the Very Large Array (VLA) at multiple wavelengths between 7 mm and 15 cm (41 GHz down to 2 GHz), as well as single-epoch Atacama Large Millimeter Array (ALMA) continuum observations covering the range from 0.4 to 1.3 mm (700 GHz down to 230 GHz). These observations all have sufficient angular resolution (typically better than \msec{0}{4}) to resolve the various compact sources known to exist in the system. 

The new VLA observations presented here, combined with previously published observations dating back to the late 1980s, were used to follow the proper motions of the different sources in the system. This confirms that the two sources known as A2 and B move on the plane of the sky with nearly identical velocities, tracing the overall relative motion between the Sun and the parent molecular cloud of \iras, Lynds~1689N. The sources A2$\alpha$ and A2$\beta$, previously identified with recent ejecta from the protostellar object A2 are observed to symmetrically move away from A2, and one of them (A2$\alpha$) shows some evidence of diluting into its surroundings. Somewhat unexpectedly, the position angle between A1 and A2, which had been observed to increase steadily during the period from the late 1980s to the early 2000s, is now found to have started to decrease. This might indicate that A1 corresponds to the impact of a precessing jet onto the surrounding material, or that A1/A2 are two members of a tight hierarchical triple system. 

Combining our new observations with data taken from the literature, we refined the determination of the spectrum of both component A and B. As expected for a protostar with an active outflow, the spectrum of component A changes from a shallow power law ($S_\nu \propto \nu^{0.44}$) at centimeter wavelengths to a steeper one ($S_\nu \propto \nu^{2.5}$) at millimeter wavelengths. In contrast, the spectrum of component B can be described by a single power law ($S_\nu \propto \nu^{2.28}$) over the entire range from 2 to 700 GHz (10 cm down to 0.5 mm). This suggests that emission from component B is entirely dominated by dust even at $\lambda$ = 10 cm, and that it drives no detectable outflow. We also find that the size of source B increases with frequency up to 41 GHz, remaining roughly constant (at \msec{0}{39} $\equiv$ 55 AU) at higher frequencies. We interpret this as evidence that source B is a dusty structure of finite size that becomes increasingly optically thick at higher frequencies. Finally, we find that the kinetic temperature at the center of source B is at least $\sim$ 1,000 K.

\acknowledgments A.H-G., L.L., L.F.R., and L.A.Z.\ acknowledge the support of DGAPA, UNAM (project IN112417), and of CONACyT (M\'exico).  A.H-G., E.C., L.L and S.B. acknowledge the financial support of the France/Mexico CONACyT -- ECOS-Nord Project N$^{\circ}$ M14U01, SPECIMEN : {\bf S}tructure {\bf P}hysiqu{\bf E} et {\bf CI}n\'{e}{\bf M}atiqu{\bf E} d'IRAS 16293 : mol\'{e}cules et conti{\bf N}uum. D.Q. acknowledges the financial support received from the STFC through an Ernest Rutherford Grant and Fellowship (proposal number ST/M004139). This paper makes use of the following ALMA data: ADS/JAO.ALMA\#2013.1.00061.S,
ADS/JAO.AL-
\noindent MA\#2013.1.00018.S and ADS/JAO.ALMA\#2011.
\noindent 0.00007.SV. ALMA is a partnership of ESO (representing its member states), NSF (USA) and NINS (Japan), together with NRC (Canada), MOST and ASIAA (Taiwan), and KASI (Republic of Korea), in cooperation with the Republic of Chile. The Joint ALMA Observatory is operated by ESO, AUI/NRAO and NAOJ. The National Radio Astronomy Observatory is a facility of the National Science Foundation operated under cooperative agreement by Associated Universities, Inc.

\bibliography{paper}{}

\end{document}